\documentclass{amsproc}
\usepackage{amssymb}

\theoremstyle{definition}

\theoremstyle{remark}

\numberwithin{equation}{section}

\newcommand{\be}{\begin{equation}}
\newcommand{\ee}{\end{equation}}
\newcommand{\bea}{\begin{eqnarray}}
\newcommand{\eea}{\end{eqnarray}}
\newcommand{\IR}{\mathbb{R}}
\newcommand{\IC}{\mathbb{C}}
\newcommand{\IZ}{\mathbb{Z}}

\newcommand{\cF}{\mathcal{F}}
\newcommand{\cV}{\mathcal{V}}
\newcommand{\cA}{\mathcal{A}}

\newcommand{\cS}{\mathcal{S}}

\newcommand{\cH}{\mathcal{H}}

\newcommand{\cM}{\mathcal{M}}

\newcommand{\cN}{\mathcal{N}}
\newcommand{\cE}{\mathcal{E}}

\newcommand{\cO}{\mathcal{O}}
\newcommand{\cR}{\mathcal{R}}
\newcommand{\cT}{\mathcal{T}}

\newcommand{\cP}{\mathcal{P}}

\newcommand{\de}{\mathrm{d}}

\newcommand{\I}{\mathrm{i}}
\newcommand{\zetastar}{\zeta^\star}
\newcommand{\Rstar}{\mathcal{R}^\star}

\newcommand{\Tr}{{\rm Tr}}

\begin{document}

\title{Rankin-Selberg methods for closed string amplitudes}

\author{Boris Pioline}
\address{CERN, Theory Unit, 
PH-TH,
Case C01600,
CH-1211 Geneva 23. }
\address{On leave from: Sorbonne Universit\'es, UPMC Univ Paris 06, UMR 7589, LPTHE, F-75005, Paris, France}
\address{On leave from: CNRS, UMR 7589, LPTHE, F-75005, Paris, France}

\email{boris.pioline@cern.ch}
\thanks{Contribution to the proceedings of the String-Math conference, June 17-21, 2013, Simons Center for Geometry of Physics, Stony Brook. \hfill CERN-PH-TH/2014-007, arXiv:1401.4265v2}

%    General info
%\subjclass{Primary 54C40, 14E20; Secondary 46E25, 20C20}

\date{October 11, 2013}

%\keywords{Differential geometry, algebraic geometry}

\begin{abstract}
After integrating over supermoduli and vertex operator positions, scattering amplitudes in superstring theory at genus $h\leq 3$ are reduced to an integral of a Siegel modular function 
of degree $h$ on a fundamental domain of the Siegel upper half plane. A direct computation
is in general unwieldy, but becomes feasible if the integrand can be expressed as a sum over
images under a suitable subgroup of the Siegel modular group: if so, the integration domain can be extended to a simpler domain at the expense of keeping a single term in each orbit -- a
technique known as the Rankin-Selberg method. Motivated by applications to BPS-saturated amplitudes,  Angelantonj, Florakis and I have applied this technique to one-loop modular integrals 
where the integrand is the product of a Siegel-Narain theta function times a weakly, almost holomorphic modular form. I survey our main results, and take some steps in 
extending this method to  genus greater than one. 
\end{abstract}

\maketitle

\section{Introduction}

According to the basic postulate of superstring theory,  scattering amplitudes of $n$
external states at $h$-th order in perturbation theory are given by correlation functions of $n$
vertex operators, integrated over the  moduli space $\mathfrak{M}_{h,n}$ of $n$-punctured super-Riemann surfaces $\Sigma$ of genus $h$ \cite{D'Hoker:1988ta,D'Hoker:2002gw}. After integrating over the fermionic moduli and the locations of the punctures,\footnote{For $h>2$
 there is no canonical way of performing these integrations,  due in part
to the non-projectiveness of supermoduli space  \cite{Witten:2012bh,Donagi:2013dua}, but
different prescriptions are expected to lead to the same integrand on $\cM_{h,0}$, up to
total derivatives.} such amplitudes reduce to an integral over the moduli space $\cM_{h,0}$ of ordinary Riemann surfaces without marked point. The latter is a quotient of the Teichm\"uller space $\cT_h$ by the mapping class group $\Gamma_h$. For $1\leq h\leq 3$, $\cT_h$ is isomorphic (via the period map, $\Sigma\mapsto \Omega\equiv \Omega_1+\I\Omega_2$, away
from suitable divisors) to the degree $h$ Siegel upper-half plane $\cH_h$ , while $\Gamma_h$ is identified with the Siegel modular group $Sp(h,\IZ)$, acting on $\cH_h$ by 
fractional linear transformations. Thus, scattering amplitudes at genus $h$ are ultimately written
as modular integrals
\be
\label{modinth}
\cA_h= {\rm R.N.}\,\int_{\cF_h}\, \de\mu_h\, F_h(\Omega)\ ,\quad
\de\mu_h = |\Omega_2|^{-\tfrac{h+1}{2}}\, 
\de \Omega_1 \, \de\Omega_2
\ee
where $\cF_h=\Gamma_h\backslash \cH_h$ is a fundamental domain of the action of $\Gamma_h$ on $\cH_h$,
$F(\Omega)$ is a  function on $\cH_h$ invariant under $\Gamma_h$, 
$|\Omega_2|=\det[\Im(\Omega)]>0$, $\de\mu_h$ is the standard invariant measure on $\cH_h$, 
and ${\rm R.N.}$ denotes a suitable infrared renormalization prescription.
For $h>3$, the Teichm\"uller space $\cT_h$ embeds as a subvariety of codimension one or greater in the Siegel upper-half plane
$\cH_h$, and $\de\mu_h$ is replaced by a suitable measure with support on $\cF_h$ -- a
complication which we shall not confront.

In general,  integrals of the type \eqref{modinth} are untractable (except, perhaps, 
numerically), due to the complicated nature of the modular function $F_h$, but also  to the unwieldy shape of any fundamental domain: for $h=2$, it takes no 
less than 25 inequalities to define $\cF_2$ ! \cite{zbMATH03144647} Yet, the computation of such integrals is an unavoidable step in extracting any phenomenological prediction of  superstring theory, and in investigating some of its structural properties such as invariance under dualities. The 
daunting task of integrating \eqref{modinth}, however,
can be considerably simplified if the integrand  can be written as a sum
over images (or Poincar\'e series) of a given function $f_h(\Omega)$,
\be
F_h(\Omega) = \sum_{\gamma\in \Gamma_{h,\infty} \backslash\Gamma_h} f_h\vert_{\gamma} (\Omega)
\ee
where $f_h\vert_{\gamma}(\Omega) = f_h(\gamma\cdot \Omega)$ and $f_h(\Omega)$ is invariant under a subgroup $\Gamma_{h,\infty}\subset \Gamma_h$. If the sum is absolutely convergent, 
exchanging it with the integral extends the integration domain to a larger fundamental domain 
$\cF_{h,\infty}=\Gamma_{\infty,h}\backslash \cH_h$, while restricting the sum to a single coset,
\be
\label{modinthu}
\cA_h= {\rm R.N.}\, \int_{\cF_{\infty,h}}\, \de\mu_h\, f_h(\Omega)\ .
\ee
This `unfolding trick', at the heart of the Rankin-Selberg method in number theory, is 
expedient if both $\cF_{h,\infty}$ and $f_h$ are simpler than $\cF_h$
and $F_h$.

A special class of amplitudes where this method is advantageous arises when the integrand $F_h(\Omega)$ factorizes as $\Phi(\Omega)\times \varGamma_{d+k,d,h}$,
where $\Phi(\Omega)$ is a (non-holomorphic, in general) 
Siegel modular form of weight $-k/2$ and $\varGamma_{d+k,d,h}$
is the Siegel-Narain theta series 
\be
\label{defSiegel}
 \varGamma_{d+k,d,h}(G,B,Y;\Omega) = |\Omega_2|^{d/2}
{\sum\limits_{p^\alpha\in\Lambda}} e^{-\pi \Omega_{2,\alpha\beta} \cM^2(p^\alpha,p^\beta)
+2\pi\I \Omega_{1,\alpha\beta} \langle p^\alpha, p^\beta \rangle} 
\ee
where the sum runs over $h$-tuples of vectors $p^\alpha$, $\alpha=1\dots h$ in an
even self-dual lattice of signature $(d+k,d)$ (hence $k\in 8\IZ$) with quadratic form $\langle \cdot, \cdot \rangle\in 2\IZ$ 
equipped with positive definite quadratic form $\cM^2(\cdot,\cdot)$. The latter
is parametrized by the Grassmannian
\be
\label{defgrass}
G_{d+k,d} = [SO(d+k)\times SO(d)] \backslash SO(d+k,d)\ ,
\ee
which can be coordinatized by a real positive definite symmetric matrix $G_{ij}$, a real antisymmetric matrix $B_{ij}$ and a real rectangular matrix $Y_{i,a}$ ($i,j=1\dots d, a=1\dots k$).
This type of theta series  arises in any string vacua which involve a $d$-dimensional 
torus with constant metric $G_{ij}$ and Kalb-Ramond field $B_{ij}$, equipped with a $U(1)^k$
flat connection with holonomies $Y_{i,a}$. Importantly, $\varGamma_{d+k,d,h}$ is invariant
under $\Gamma_h\times O(d+k,d,\IZ)$, where the last factor is  the
automorphism group  of the lattice $\Lambda$ (also known as T-duality group), acting by right-multiplication on the coset \eqref{defgrass}. 

In such cases, the integral 
\be
\label{modinth2}
\cA_h= {\rm R.N.}\, \int_{\cF_h}\, \de\mu_h\, \Gamma_{d+k,d,h}(G,B,Y;\Omega)\, \Phi(\Omega)\ .
\ee 
can be computed by  expressing  $\Gamma_{d+k,d,h}$ as a sum of Poincar\'e series 
under $\Gamma_h$,
and applying the unfolding trick to each one of them. This `lattice unfolding technique' has been the method of choice for one-loop amplitudes in the physics literature  \cite{McClain:1986id,O'Brien:1987pn,Dixon:1990pc,Mayr:1993mq,Harvey:1995fq,Harvey:1996gc, Bachas:1997mc, Lerche:1998nx, Foerger:1998kw, 
Kiritsis:1997hf, Kiritsis:1997em, Marino:1998pg}, and has been very useful in extracting asymptotic expansions at
particular boundary components of $G_{d+k,d}$. However, its main drawback 
is that the  various terms in the Poincar\'e series decomposition are not invariant under  
$SO(d+k,d,\IZ)$, even though the sum is. As a result, the result of the unfolding trick is not
manifestly invariant under $SO(d+k,d,\IZ)$.

Another option, advocated in  \cite{Angelantonj:2011br}\footnote{The idea of applying Rankin-Selberg-type methods to compute one-loop modular integrals, albeit in a rather different way from  \cite{Angelantonj:2011br}, was first put forward in \cite{Cardella:2008nz}. Steps
towards extending them to higher genus were taken in \cite{Cacciatori:2011qd}.}
 and further developed in  \cite{Angelantonj:2012gw,Angelantonj:2013eja} (see \cite{Florakis:2013ura} for a complementary survey), is to represent the other factor $\Phi(\Omega)$  in the integrand (or the full integrand, in the absence of any Siegel-Narain theta series) as a Poincar\'e series, and use {\it it} to unfold the integration  domain. This technique is of course limited by our ability to find absolutely convergent Poincar\'e series representations for (in general, non-holomorphic) Siegel modular forms. For genus one,  it turns out that any almost, weakly holomorphic modular form of negative weight under $\Gamma_1=SL(2,\IZ)$ (or congruence subgroups thereof) can be represented as a linear combination of certain absolutely convergent Poincar\'e series, first introduced by Niebur \cite{0288.10010} and Hejhal  \cite{0543.10020} and revived in recent mathematical 
work  \cite{1004.11021,BruinierOno,zbMATH02135351,1154.11015}. 
Almost, weakly holomorphic integrands $\Phi(\Omega)$ are non-generic, but do occur for certain classes of 'BPS-saturated' amplitudes, which play a central role for determining threshold corrections to gauge couplings and for testing non-perturbative dualities (see e.g. \cite{Kiritsis:1999ss}
for a review).  
As we shall see, the unfolding trick produces a sum over lattice vectors with fixed
integer norm, manifestly invariant under T-duality. Physically, it can be interpreted as a sum
of  field-theory type amplitudes, with BPS states running in the loop.  In particular, it exposes the singularities of the amplitude, originating from BPS states becoming massless. The price to pay is that the behavior at the boundary components is obscured, although it can be recovered in some cases, showing agreement with -- and uncovering hidden structure in -- the result of the usual lattice unfolding technique. 

A third, and most radical option,  is to represent $1$ as a Poincar\'e series, and use {\it it} to unfold the integral. Indeed, it is well-known that $1$ is a residue of non-holomorphic Siegel-Eisenstein series, the simplest conceivable example of Poincar\'e series. This trick, which we refer to as the 
Rankin-Selberg-Zagier method, is very useful to evaluate integrals of the type \eqref{modinth2} with $\Phi=1$ (hence $k=0$). It expresses the result, for any genus, as a Langlands-Eisenstein series of $SO(d,d,\IZ)$, verifying a conjecture put forward
in \cite{Obers:1999um}.

We begin our survey of various applications of the Rankin-Selberg method (or unfolding trick)
to closed string amplitudes in \S\ref{sec_1looptriv} by computing one-loop integrals of symmetric lattice partition functions by means of a non-holomorphic Eisenstein series insertion. In \S\ref{sec_1loopgen} we move on to more general 
modular integrals of non-symmetric lattice partition functions against harmonic elliptic genus,
which we compute by representing the latter as linear combination of Niebur-Poincar\'e series.
In the last section \S\ref{sec_highergen} we take steps towards extending the Rankin-Selberg-Zagier method  to higher genus. 

\medskip 

{\noindent \it Acknowledgements:} I wish to thank C. Angelantonj and I. Florakis for a very
enjoyable collaboration on the results reported in \S\ref{sec_1looptriv}-\ref{sec_1loopgen}, K. Bringmann and D. Zagier for valuable advice during the course of this project, R. Donagi
and R. Russo for comments on an earlier version of this manuscript, 
and the organizers of the String Math 2013 conference for their kind invitation to speak.

%%%%%%%%%%%%%%%%%%

\section{One-loop modular integrals with trivial elliptic genus \label{sec_1looptriv}}

We start with the simplest case of one-loop modular integrals of the form
\eqref{modinth2} with $\Phi=1$ (hence $k=0$).  Such integrals
were computed by the `lattice unfolding technique' in \cite{McClain:1986id,O'Brien:1987pn} for $d=1$, \cite{Dixon:1990pc} for $d=2$, \cite{Kiritsis:1997em} for $d\geq 3$, and a conjectural
relation to constrained Epstein series was put forward in \cite{Obers:1999um}. In this section,
we shall calculate them instead by inserting by hand a non-holomorphic Eisenstein series 
in the integrand, computing the integral by the unfolding trick and taking a suitable residue  at the end. We use the standard notations $\tau=\tau_1+\I\tau_2$,
$q=e^{2\pi \I\tau}$ for the period $\Omega_{11}$ and modulus of an elliptic curve, and write
$\Gamma=\Gamma_1=SL(2,\IZ), \cH_1=\cH$, etc.

\subsection{Non-holomorphic Eisenstein series \label{sec_eis}}

The non-holomorphic Eisenstein series for $\Gamma=SL(2,\IZ)$ is defined by
the sum over images
\be
\label{Edefinition}
\begin{split}
 E^\star(s;\tau) = & \zetastar (2s) 
 \sum_{\gamma \in \varGamma_\infty \backslash \varGamma}
 \tau_2^s \vert \gamma 
 =  \tfrac{1}{2}\,  \zetastar (2s)\, \sum_{(c,d)=1}  
      \frac{\tau_2^s }{|c\, \tau + d|^{2s}}\ ,
\end{split}
\ee
where $\Gamma_{\infty}$ is the subgroup of upper-triangular matrices in $\Gamma$, and
 \be
 \zetastar (s) \equiv \pi^{-s/2}\, \varGamma (s/2) \, \zeta (s) = \zetastar(1-s)
 \ee
is the completed Riemann zeta function. The sum  converges absolutely for $\Re(s)>1$, and has a meromorphic continuation to all $s$. The normalization in \eqref{Edefinition} ensures
that $E^\star(s;\tau)$ invariant under $s\mapsto  1-s$, and has only simple poles at  $s=0$ and $s=1$. The key point for our purposes is that the residue at $s=1$ is constant -- in agreement with the fact that $E^\star(s;\tau)$ is an eigenmode of the Laplacian $\Delta_{\cH}$ on $\cH$, with vanishing
eigenvalue at $s=0$ or 1,
\be
\left[ \Delta_{\cH} -\tfrac{1}{2}\, s(s-1) \right]\, E^\star (s;\tau) = 0\  ,\quad 
 \Delta_{\cH} = 2\tau_2^2\, \partial_{\tau} \partial_{\bar\tau}\ .
\ee
More precisely, the Laurent expansion at $s=1$ is given by the first Kronecker limit formula, 
\be
\label{kron1}
E^\star(s;\tau) =\frac{1}{2(s-1)} + \tfrac{1}{2} \left( \gamma_E -  \log ( 4 \pi \, \tau_2 \, |\eta(\tau)|^4) \right)
+ \cO(s-1) \ ,
\ee
where $\eta=q^{1/24}\prod_{n\geq 1}(1-q^n)$ is the Dedekind eta function and $\gamma_E$
is Euler's constant. All these statements
are easy consequences of the Fourier series representation of $E^\star(s;\tau)$, or Chowla-Selberg formula,
  \be
  \begin{split}
 \label{chowla}
 E^\star(s;\tau) =& \zetastar(2s) \, \tau_2^s + \zetastar(2s-1) \, \tau_2^{1-s} \\
 &
 + 2 \, \sum_{N\neq 0} |N|^{s-\frac12} \sigma_{1-2s}(N) \, \tau_2^{1/2} K_{s-\frac12}(2\pi |N| \tau_2)\,
 e^{2\pi \I N \tau_1}\,,
 \end{split}
 \ee
 where  $\sigma_t(N)=\sum_{d|N} d^t$ is the divisor function  and $K_t(z)$ is the modified Bessel function of the second kind. Below we shall denote the first line of \eqref{chowla}, which 
 dominates the behavior at $\tau_2\to\infty$, by $E_0^\star(s;\tau)$.
 
For any modular function $F(\tau)$ of rapid decay (such as the modulus square $|\psi|^2$
of a holomorphic cusp form), we consider the modular integral (also known as the
Rankin-Selberg transform)
\be
\label{defRstar0}
\Rstar(F;s) = \int_{\cF} \de\mu\, E^\star(s; \tau)\, F(\tau)\ .
\ee
For $\Re(s)>1$, the sum over cosets in \eqref{Edefinition} can be exchanged with the integral,
so that the integration domain is extended to the strip 
\be
\cS=\Gamma_{\infty} \backslash \Gamma = \{\tau_2>0, -\tfrac12<\tau_1\leq \tfrac12\}\ ,
\ee
at the expense of retaining the contribution
of the unit coset only,
\be
\label{RSunfold}
\begin{split}
\Rstar(F;s) =
\zetastar (2s)\, \int_\cS \de\mu\, \tau_2^s\, F(\tau)
=\zetastar(2s)\, \int_0^\infty \de\tau_2\, \tau_2^{s-2}\, F_0(\tau_2) \ .
\end{split}
\ee
The last equality expresses $\Rstar(F;s)$ as a Mellin transform of the zero-th Fourier
coefficient $F_0(\tau_2)=\int_{-1/2}^{1/2}  \de\tau_1 \, F(\tau)$. At the same time, $\Rstar(F;s)$
inherits the  meromorphicity and invariance under $s\mapsto 1-s$ satisfied by $E^\star$.  
In the case where $F=|\psi|^2$, 
$\Rstar(F;s)$ is proportional to the $L$-series $\sum |a_n|^2 n^{-s}$, whose analyticity 
and functional equation are thereby determined. This is one of the main uses of the
Rankin-Selberg method in number theory \cite{MR993311}.

More importantly for our purposes, the fact that the residue of  $E^\star(s;\tau)$ at $s=1$ is constant implies that the residue of $\Rstar(F; s)$ at $s=1$ is proportional to the  modular integral of $F$,
\be
\label{ResR}
\mathrm{Res}_{s=1}  \Rstar(F; s) 
= \tfrac{1}{2} \int_\cF \de\mu\, F  \ .%=  -\mathrm{Res}_{s=0} \, \Rstar(F; s) 
\ee
Unfortunately, this statement only holds for functions $F$ of rapid decay, which rules out 
the interesting case $F=\Gamma_{d,d,1}$.
 In the next section, following \cite{MR656029} we discuss how the unfolding
trick can nevertheless be used after proper regularization.

\subsection{Rankin-Selberg-Zagier method \label{sec_rsz}}

Let us now consider a modular function $F$ with polynomial growth at the cusp,
\be
F(\tau)\sim \varphi(\tau_2)\ ,\quad \varphi(\tau_2) = \sum_\alpha c_\alpha \tau_2^{\alpha}\ .
\ee
In order to regulate infrared divergences in the integral \eqref{defRstar0}, we truncate
the integration domain to $\cF_\cT=\cF \cap \{\tau_2>\cT\}$ and define the renormalized integral as
\be
\label{renormint}
{\rm R.N.}\, \int_{\mathcal F} \de\mu\,  F(\tau)
=\lim_{\cT\to \infty} \left[ \int_{{\mathcal F}_{\cT}} \de\mu \, F(\tau )  - \hat \varphi (\cT) \right] \ ,
\ee
where $\hat\varphi(\cT)$ is the anti-derivative of $\varphi(\tau_2)$,
\be
 \hat \varphi (\cT) = \sum_{\alpha \neq 1} c_\alpha \frac{\cT^{\alpha-1}}{\alpha-1}
 +c_1\, \log \cT\ .
\ee
On the other hand, we define the Rankin-Selberg transform of $F$ as the Mellin transform of the zero-th Fourier coefficient, minus its leading behavior, 
\be
\label{defRstar}
\Rstar(F;s) = \zetastar(2s)\, \int_0^\infty \de \tau_2\, \tau_2^{s-2}\,\left(F_0 - \varphi \right)  \ .
\ee
The renormalized integral \eqref{renormint}
is then related to the Rankin-Selberg transform \eqref{defRstar} via a generalization 
of \eqref{ResR} \cite{MR656029}
\be
\label{ResRgen}
{\rm R.N.}\, \int_{\mathcal F} \de\mu\,  F(\tau) = 2\, {\rm Res}_{s=1}\, 
\left[ \Rstar(F;s) + \zetastar(2s) 
h_{\cT}(s) + \zetastar(2s-1) h_{\cT}(1-s) \right] - \hat \varphi (\cT) \ ,
\ee
where $h_\cT(s)$ is the meromorphic function of $s$ defined by 
\be
h_{\cT}(s) = \int_0^{\cT} \de\tau_2\, \varphi(\tau_2)\, \tau_2^{s-2} = \sum_{\alpha}  c_\alpha\, 
\frac{\cT^{\alpha+s-1}}{\alpha+s-1}\ .
\ee
Note that the right-hand side of \eqref{ResRgen} is by construction independent of the infrared cut-off $\cT$. The derivation  of \eqref{ResRgen} is based on the generalized unfolding trick for 
modular integrals on the truncated fundamental domain, 
\be
\label{gentrick}
\int_{{\mathcal F}_{\cT}} \de\mu \, F\, \sum_{\gamma\in\Gamma_{\infty}\backslash \Gamma}
f\vert_\gamma = \int_{{\mathcal S}_{\cT}} \de\mu \, F\, f -
\int_{\cF-{\mathcal F}_{\cT}} \de\mu \, F\, 
\sum_{\gamma\in\Gamma_{\infty}\backslash \Gamma\atop
\gamma\neq 1}
f\vert_\gamma \ ,
\ee
where $\cS_\cT$ is the truncated strip $\{\tau_2<\cT,-1/2<\tau_1<1/2\}$. 
Applying this observation to the regulated integral $R_{\cT}^\star(F;s) \equiv \int_{\cF_\cT} \de\mu\,
F\, E^\star(s;\tau)$ and reorganizing terms gives \cite[Eq. (27)]{MR656029}
\be
\begin{split}
\label{mastereqRS}
\Rstar(F; s) =& R_{\cT}^\star(F;s) +
\int_{\cF-\cF_{\cT}} \de\mu \, \left( F\, E^\star(s;\tau) - \varphi\, E^{\star}_0(s;\tau_2) \right) 
\\
&-\zetastar(2s) \, h_{\cT}(s) - \zetastar(2s-1)\, h_{\cT}(1-s) \ ,
\end{split}
\ee
from which \eqref{ResRgen} follows. Another consequence of \eqref{mastereqRS} is that
the Rankin-Selberg transform $\Rstar(F; s)$ has a meromorphic continuation in $s$, invariant
under $s\mapsto 1-s$, and analytic away from 
$s=0,1,\alpha_i,1-\alpha_i$. A particularly
pleasant feature of the renormalization prescription \eqref{renormint} is that the Rankin-Selberg transform $\Rstar(F; s)$ coincides
with the renormalized integral 
\be
\Rstar(F;s) = {\rm R.N.}\, \int_{\mathcal F} \de\mu\,  F(\tau)\, E^\star(s;\tau)\ .
\ee
Moreover, if $F$ is constant, $\Rstar(F;s)$ vanishes and therefore ${\rm R.N.}\, \int_{\mathcal F} \de\mu \, E^\star(s;\tau) = 0$.

\subsection{Constrained Epstein series}

The Rankin-Selberg-Zagier method discussed in the previous subsection applies immediately
to modular integrals of Siegel-Narain theta series for even self-dual lattices of signature $(d,d)$, 
\be
\label{gdd}
\varGamma_{d,d} (G,B;\tau) = 
\tau_2^{d/2} \, \sum_{(m_i,n^i)\in \IZ^{2d}} e^{-\pi \tau_2\, {\mathcal M}^2(m_i,n^i)+2\pi\I \tau_1 \, m _i \,  n^i}\ ,
\ee
where $\cM^2$ is the positive definite quadratic form
\be
\label{M2GB}
{\mathcal M}^2(m_i,n^i)= (m_i+B_{ik} n^k) G^{ij} (m_j+B_{jl} n^l) + n^i G_{ij} n^j
\ee
and $(G_{ij},B_{ij})=G_{ji},-B_{ji})$ parametrize the Grassmannian $G_{d,d}$. Eq. \eqref{gdd}
defines a modular function $F$ on $\cH$ of polynomial growth characterized by 
\be
\varphi(\tau_2) = \tau_2^{\tfrac{d}{2}}\ ,\quad
 h_{\cT} (s) = \frac{\cT^{s+\tfrac{d}{2}-1}}{s+\tfrac{d}{2} -1} \,,
\quad 
\hat\varphi (\tau_2 ) = 
\begin{cases}
\tau_2^{\tfrac{d}{2}-1}/( \tfrac{d}{2}-1) &
\mbox{if}\ d\neq 2 \\
\log\, \tau_2 & \mbox{if}\ d=2
\end{cases}\ .
\ee
Its Rankin-Selberg transform is
\be
\label{REps}
\begin{split}
\Rstar (\varGamma_{d,d};s ) 
&= \zetastar (2s) \, \int_0^\infty  \de\tau_2\, 
\tau_2^{s+\tfrac{d}{2}-2} \, \sum_{\substack{(m_i,n^i)\in\IZ^{2d}\backslash(0,0)\\ m_i n^i=0}}   
e^{-\pi \tau_2\cM^2(m_i,n^i)} \\
&=
%\zetastar (2s) \, \frac{\varGamma (s+\tfrac{d}{2}-1)}{\pi^{s+\tfrac{d}{2}-1}} \, 
%\mathcal{E}^d_V (s+\tfrac{d}{2}-1;G,B)  \equiv 
 {\mathcal E}^{SO(d,d),\star}_V (s+\tfrac{d}{2}-1;G,B)  
\end{split}
\ee
where 
$\mathcal{E}^{SO(d,d),\star}_V(s)$ is the {\it completed, constrained Epstein series}
defined by \cite{Obers:1999um}
\be
\label{defceps}
\begin{split}
\mathcal{E}^{SO(d,d),\star}_V(s) =& \zetastar (2s) \, \zetastar (2s+2-d) \,
\mathcal{E}^{SO(d,d)}_V(s)
\\
\mathcal{E}^{SO(d,d)}_V (s;G,B) =&\frac{1}{\zeta(2s)} \sum_{\substack{(m_i,n^i)\in\IZ^{2d}\backslash(0,0)\\ m_i n^i=0}} [{\mathcal M}^2(m_i,n^i)]^{-s}\ ,
\end{split}
\ee
the sum being absolutely
convergent for $\Re(s)>d$.
The results of \S\ref{sec_rsz} show that ${\mathcal E}^{SO(d,d),\star}_V (s)$ admits 
 a meromorphic continuation in $s$, invariant under $s\mapsto d-1-s$, 
 with simple poles at $s=0, \frac{d}{2}-1, \frac{d}{2}, d-1$ (or
 double poles at $s=0$ and $s=1$ if $d=2$). For $d\neq 2$, the residues at $s=\tfrac{d}{2}$ 
 or $s=\tfrac{d}{2}-1$ produce the modular integral of interest:
\be\label{Idb2}
\begin{split}
 {\rm R.N.}\, \int_{\mathcal F} \de\mu \, \varGamma_{d,d}  &= 
2\, {\rm Res}_{s=\tfrac{d}{2}}  {\mathcal E}^{SO(d,d),\star}_V \left(s; G,B\right)  
 = \zetastar(d-2)\, {\mathcal E}^{SO(d,d)}_V \left(\tfrac{d}{2}-1; G,B\right)\ ,
\end{split}
\ee
rigorously proving a conjecture in \cite{Obers:1999um}. Physically, the integral \eqref{Idb2} computes (among other things) the one-loop contribution to $\cR^4$ couplings in type II strings compactified on a torus $T^d$ \cite{Green:1997di,Kiritsis:1997em}. The
right-hand side is interpreted as
a sum of one-loop contributions from particles of momentum $m_i$ and winding $n^i$ along
the torus, with mass $\cM^2(m_i,n^i)$, satisfying the BPS constraint $m_in^i=0$. It is manifestly
invariant under the T-duality group $SO(d,d,\IZ)$, under which $m_i,n^i$ transform in the vector
(defining) representation. For $s\to 1$ the $s$-dependent generalization \eqref{REps} can be thought of
as the  dimensionally regularized\footnote{Other versions of dimensional
regularization in string theory were discussed in \cite{Green:1982sw,Bern:1991aq,Kiritsis:1994ta}.} amplitude, i.e. the amplitude in $D=10-(d+2s-2)$ non-compact
dimensions. The case $s=2$ also computes $D^4 \cR^4$ couplings at one-loop in type II string theory on $T^d$ \cite{Green:1999pv}. Mathematically, \eqref{defceps} is recognized as the degenerate Langlands-Eisenstein series of $SO(d,d)$ with infinitesimal character $\rho-2s\lambda_1$ (where $\rho$ is the Weyl vector and $\lambda_1$ the weight of the vector 
representation) \cite{Green:2010wi,Green:2010kv}.  The residue of this
Langlands-Eisenstein series at $s=\tfrac{d}{2}$
yields the minimal theta series associated to the minimal representation of $SO(d,d)
$ \cite{Pioline:2010kb}. 

For $d=2$, the Grassmannian $G_{2,2}$ reduces to a product of two upper half planes $\cH_T\times
\cH_U$, where $T$ and $U$ are the K\"ahler modulus and complex structure modulus, respectively,
while the T-duality group decomposes into $SL(2,\IZ)_T\times SL(2,\IZ)_U\ltimes \sigma_{T,U}$, 
where $\sigma_{T,U}$ is an involution exchanging $T$ and $U$.
The BPS constraint $m_1 n^1+m_2 n^2=0$ can be solved explicitly, allowing to rewrite the
 constrained Epstein series as a product of two non-holomorphic Eisenstein series \cite{Angelantonj:2011br},
\be
\begin{split}
 {\mathcal E}^{SO(2,2),\star}_V (s;T,U) = & 2\,  E^\star (s;T) \, E^\star (s;U)\ .
\end{split}
\ee
Extracting the residue at $s=0$ or $1$ by means of the Kronecker limit formula \eqref{kron1} 
leads to
\be
\label{dkl}
\int_{{\mathcal F}} \varGamma_{2,2} (T,U;\tau) \, \de\mu 
= - \log \left(  T_2 \, U_2 \, |\eta (T)\, \eta (U) |^4 \right) + {\rm cte}\ ,
\ee
which agrees with \cite{Dixon:1990pc}, up to a renormalization scheme-dependent additive constant. 
The reader familiar with \cite{Dixon:1990pc} may appreciate the 
elegance of the Rankin-Selberg-Zagier method compared with the lattice unfolding method.

%%%%%%%%%%%%%%%%%%

\section{One-loop modular integrals with harmonic elliptic genus \label{sec_1loopgen}}

While the integrand $\Phi$ in type II one-loop string amplitudes is of polynomial 
growth at the
cusp $\tau\to\I\infty$, this is not the case for heterotic strings, due to  the tachyon in the spectrum before 
imposing the GSO projection. Instead,  $\Phi$ is a modular form of negative modular weight
$w=-k/2$ with a first order pole at the cusp. The Rankin-Selberg-Zagier method described in \S\ref{sec_rsz} is not directly applicable, however  the unfolding trick could still be used 
provided $\Phi$ had  a uniformly convergent
Poincar\'e series representation
\be 
\label{poincaintro}
 \Phi = \sum_{\gamma \in \varGamma_\infty \backslash \varGamma}  f\vert_{w}\gamma
\ee
with suitable $\Gamma_{\infty}$-invariant seed $f(\tau)$. $f$ should grow as $1/q^\kappa$ at $\tau_2\to\infty$ if \eqref{poincaintro} is to represent a modular form with a $\kappa$-th order
pole at the cusp, but uniform convergence requires $f(\tau) \ll \tau_2^{1-\frac{w}{2}}$ as $\tau_2\to 0$. The naive choice $f(\tau)=1/q^\kappa$ is fine for weight $w>2$ but fails for $w\leq 2$.

\subsection{Selberg-Poincar\'e and Niebur-Poincar\'e series}

A first, natural option for regulating the sum is to insert a non-holomorphic convergence factor \`a la 
 Hecke-Kronecker,  i.e. choose  a seed $f(\tau)=\tau_2^{s-\frac{w}{2}}\, q^{-\kappa}$. The resulting
{\it Selberg-Poincar\'e series} 
\be
\label{Eskw}
E (s,\kappa ,w;\tau)\equiv  \tfrac{1}{2} \sum_{(c,d)=1} 
\frac{(c\tau+d)^{-w}\, \tau_2^{s-\frac{w}{2}}}{|c\tau+d|^{2s-w}}\, 
e^{-2\pi\I\kappa\, \frac{a\tau+b}{c\tau+d}}\, 
\ee
converges absolutely for $\Re(s)>1$, but analytic continuation to the desired value
$s=\tfrac{w}{2}$ is non-trivial, as it depends on deep properties of Kloosterman sums 
is tricky, and in general non-holomorphic \cite{0142.33903,0507.10029}. Another undesirable
feature of the Selberg-Poincar\'e series \eqref{Eskw} is that it is not an eigenmode of the 
weight $w$ Laplacian on $\cH$, rather\footnote{Here $\Delta_{\cH,w}=2D_{w-2}\bar D_w$ where 
$D_w$, $\bar D_w$ are defined in \eqref{modularderiv}. Notice the change of convention compared to \cite{Angelantonj:2012gw}.}
\be\label{laplEskw}
\left[\Delta_{\cH,w} + \frac12 (s-\tfrac{w}{2})(1-\tfrac{w}{2}-s) \right] \, E(s,\kappa,w) = 
2\pi\kappa\, ({s-\tfrac{w}{2})}\, 
E(s+1,\kappa,w)\ ,
\ee
so that the analytic continuation to $s=\tfrac{w}{2}$ is not guaranteed to yield a holomorphic
result, nor even harmonic \cite{Pribitkin09}. 

A much more convenient choice,   which does not require analytic continuation, is
 the {\it Niebur-Poincar\'e series}\footnote{The relation between the Poincar\'e series
 \eqref{Eskw} and \eqref{Fskw} can be found in \cite[App. B]{Angelantonj:2012gw}.}
\be
\label{Fskw}
\cF(s,\kappa,w;\tau) =\tfrac12\sum_{(c,d)=1} 
(c\tau+d)^{-w}\,
\cM_{s,w}\left(-\frac{\kappa\tau_2}{|c\tau+d|^2}\right)\, e^{-2\pi\I \kappa \Re(\frac{a\tau+b}{c\tau+d})}
\ee
first introduced by Niebur \cite{0288.10010} (for weight zero) and Hejhal  \cite{0543.10020} and revived in recent mathematical work  on Mock modular forms\cite{1004.11021,BruinierOno,zbMATH02135351, 1154.11015}. The 
seed $f(\tau)=\cM_{s,w}(-\kappa\tau_2)\, e^{-2\pi\I\kappa\tau_1}$, where 
\be
\cM_{s,w}(y) = |4\pi y|^{-\frac{w}{2}}\, M_{\frac{w}{2} {\rm sgn}(y), s-\frac12}
\left(4\pi |y| \right)\ ,
\ee
is proportional to the Whittaker function $M_{\lambda,\mu}(z)$, is uniquely determined by 
the requirements that  $\cF(s,\kappa,w;\tau)$ be an eigenmode of the Laplacian on $\cH$,
\be
\label{lapF}
\left[\Delta_{\cH,w} + \tfrac{1}{2}\, (s-\tfrac{w}{2})(1-\tfrac{w}{2}-s) \right] \, \cF(s,\kappa,w;\tau) = 0\ ,
\ee
and that $f(\tau)$ has the desired growth at $\tau_2\to\infty$ and $\tau_2\to 0$,
\be
f(\tau)\sim_{\tau_2\to \infty} 
\frac{\varGamma(2s)}{\varGamma(s+\frac{w}{2})}\, q^{-\kappa}\ ,\quad
f(\tau)\sim_{\tau_2\to 0} |4\pi\kappa\tau_2|^{s-\frac{w}{2}}e^{-2\pi\I\kappa\tau_1}\ .
\ee
The last property ensures that $\cF(s,\kappa,w)$ converges absolutely for $\Re(s)>1$,
while a more detailed argument based on the Fourier expansion of $\cF(s,\kappa,w)$ 
(which can be found in \cite{1154.11015,Angelantonj:2012gw}) shows that $\cF(s,\kappa,w)$ is holomorphic for $\Re(s)>\tfrac34$ \cite{zbMATH02121181,DukeImamogluToth}.
Besides being an eigenmode of $\Delta_{\cH,w}$,  $\cF(s,\kappa,w)$ also transforms
in a simple way under the raising, lowering and Hecke operators $D, \bar D, H_m$ defined by
\be
\label{modularderiv}
D_w=\tfrac{\I}{\pi} \left( \partial_\tau-\frac{\I w}{2\tau_2} \right) \ ,\qquad
\bar D_w=-\I \pi\, \tau_2^2 \partial_{\bar\tau}\,,
\ee
\be
\label{defHecke}
(H_m \cdot \Phi)(\tau)=\sum_{\substack{a,d>0 \\ ad=m}} \sum_{b \, {\rm mod}\, d} 
d^{-w}\, \Phi\left( \frac{a\tau+b}{d}\right)\,,
\ee
namely \cite{Angelantonj:2012gw}
\be
\label{DwF}
\begin{split}
D_w\cdot \cF(s,\kappa,w;\tau) &= 2\kappa\, (s+\tfrac{w}{2})\,   \cF(s,\kappa,w+2;\tau)\,,
\\
\bar D_w \cdot \cF(s,\kappa,w;\tau) &= \frac{1}{8\kappa} (s-\tfrac{w}{2})\,   \cF (s,\kappa,w-2;\tau)\,.
\\
H_{m}\cdot   \cF(s,\kappa,w;\tau) &=  
\sum_{d|(\kappa,m)} d^{1-w}\, \cF(s,\kappa m/d^2,w;\tau)\ .
\end{split}
\ee

The
decisive advantage of Niebur's Poincar\'e series over Selberg's, however, is that
the value $s=1-\tfrac{w}{2}$, degenerate with the value $s=\tfrac{w}{2}$ under the Laplacian
\eqref{lapF}, lies in the convergence domain $\Re(s)>1$ (except for $w=0$, which requires
a more careful treatment). Eigenmodes of the Laplacian \eqref{lapF} with $s=\tfrac{w}{2}$
or equivalently $s=1-\tfrac{w}{2}$ are known as {\it weak harmonic Maass forms} (WHMS), and have
a Fourier expansion near $\tau=\I\infty$ of the form\footnote{Here we restrict to the case where
the shadow is regular at $\tau=\I\infty$, see \cite{DukeImamogluToth} for the general expansion.} 
 \be
\label{genharm}
\Phi = \sum_{m=-\kappa}^{\infty} a_m\, q^m 
+\tfrac{\ (4\pi\tau_2)^{1-w}}{w-1}\bar b_0\,
+
\sum_{m=1}^{\infty} m^{w-1}\, 
\bar b_{m}\, \Gamma(1-w, 4\pi m\tau_2)\, q^{-m}
\ee
Weak holomorphic modular forms are a special case of WHMS,
where the negative frequency coefficients $\bar b_m$ vanish. Mock modular forms
are defined as the analytic part $\Phi^+=\sum_{m=-\kappa}^{\infty} a_m\, q^m$  of a WHMS. 
Acting on any WHMS $\Phi$ of weight $w$ with the lowering operator $\bar D$ produces the complex conjugate of a
holomorphic modular form $\Psi=\sum_{m\geq 1} b_m q^m$ of weight $2-w$ (the {\it shadow}) while the iterated raising operator $D^{1-w}$ produces a weakly holomorphic
modular form $\Xi=\sum_{m=-\kappa}^{\infty} m^{1-w} a_m\, q^m$  of weight $2-w$ (the {\it ghost} ?) such that 
 $\Phi^+$ is an Eichler integral of $\Xi$. In the case of $\cF(1-\tfrac{w}{2},\kappa,w)$, 
the shadow  is the usual Poincar\'e series $\Psi\propto P(-\kappa,2-w)=\sum q^{\kappa}\vert_{\gamma,2-w}$,
while the ghost is the Niebur-Poincar\'e series $\Xi\propto \cF(1-\tfrac{w}{2},\kappa,2-w)$. In particular, $\Psi$ is a cusp form of weight $2-w$, so must vanish for $w=0,-2,-4,-6,-8,-12$.
Indeed, for these values, $\cF(1-\tfrac{w}{2},\kappa,w)$ is an ordinary weak holomorphic
modular form, e.g. 
\be
\label{Fex}
\cF(1,1,0) = J+24\ ,\quad \cF(2,1,-2) = 3!\, \tfrac{E_4 E_6}{\Delta}\ ,\quad \cF(7,1,-12) = 13!\, /\Delta\ ,\dots
\ee
where $E_4,E_6$ are the usual Eisenstein series of weight $4,6$ under $SL(2,\IZ)$, 
$\Delta=\eta^{24}$ is the modular discriminant and  $J=\tfrac{E_4^3}{\Delta}-744=1/q+\cO(q)$ is the usual Hauptmodul.  In contrast, for $w=-10$, $\cF(1,1,-10)$ is a genuine WHMS,
with irrational positive frequency Fourier coefficients and non trivial shadow, proportional to $\Delta$ \cite{OnoMockDelta}.

It is worth noting that for  $s=1-\tfrac{w}{2}$, the seed of the Niebur-Poincar\'e series simplifies to 
\be
f(\tau) = 
\Gamma(2-w)\, \left( q^{-\kappa} - \bar q^{\kappa} \, 
\sum_{\ell=0}^{-w} \frac{(4\pi\kappa\tau_2)^\ell}{\ell !} \right)\ ,
\ee
which plainly shows  the improved ultraviolet behavior compared to the 
naive choice $f(\tau)\sim q^{-\kappa}$.  

Using the Niebur-Poincar\'e series $\cF(s,\kappa,w)$ with $s=1-\tfrac{w}{2}$, we can now
represent any weakly holomorphic modular form $\Phi$ of weight $w\leq 0$ as 
a linear combination of Niebur-Poincar\'e series\footnote{Similarly,  
modular forms under congruence subgroups of $SL(2,\IZ)$ can be represented as linear combinations
of Niebur-Poincar\'e series attached to all cusps, see \cite{Angelantonj:2013eja} for the example of the Hecke congruence subgroup $\Gamma_0(N)$.}
\be
\label{Phirep}
\Phi = \frac{1}{\Gamma(2-w)}\, 
\sum_{-\kappa\leq m<0} \, a_m \,   \cF(1-\tfrac{w}{2},-m,w;\tau) + a_0'\, \delta_{w,0}
\ee
where the coefficients are read off from 
the polar part $\Phi= \sum_{-\kappa\leq m\leq 0} \,  a_m\, q^{m} +\cO(1)$ at the 
cusp $\tau\to\I\infty$. Indeed, the difference between the left and right-hand
sides of \eqref{Phirep} is a harmonic Maass form of negative weight which is exponentially suppressed at the
cusp (for suitable choice of $a_0'$ if $w=0$), hence vanishes  \cite{1004.11021}. In particular,
while each term in \eqref{Phirep} may be a WHMF with non-trivial shadow, the shadow cancels
in the linear combination \eqref{Phirep}. More generally, using the fact that almost, weakly holomorphic
modular forms of weight $w<0$ are linear combinations\footnote{This is not the case
for almost, weakly holomorphic modular forms of weight $w>0$, $(\hat E_2)^n J$ being
a counter-example. Such cases can be treated by considering $s$-derivatives of  
$\cF(s,\kappa,w)$ at $s=w/2$ \cite{AFP4-to-appear}.
} of  iterated derivatives $D^n \Phi_{w-2n} $ of weakly holomorphic
modular forms of weight $w-2n$, along with \eqref{DwF}, we can similarly represent any almost, weakly holomorphic
modular form of weight $w<0$ as a linear combination
\be
\label{Phiexp}
\Phi = \sum_{p=0}^{n}  \sum_{-\kappa\leq m<0} a^{(p)}(m)\, \cF(1-\tfrac{w}{2}+p,-m,w;\tau)\ + a_0' \, \delta_{w,0}
\ee
where $n$ is the depth  (i.e. the maximal power of $\hat E_2$). As an example relevant
for the computation of threshold corrections to gauge couplings in heterotic string theory
compactified on $K_3\times T^2$, we quote
\be
\frac{\hat E_2 \, E_4 \, E_6 - E_6^2}{\Delta} = {\mathcal F} (2,1,0 ) - 6 \, {\mathcal F} (1,1,0) + 864\ .
\ee

\subsection{One-loop BPS state sums}

Using the representation \eqref{Phiexp}, any one-loop
modular integral of the form \eqref{modinth2} is reduced to a linear combination of
modular integrals of Niebur-Poincar\'e series against lattice-partition functions,
\be
 {\mathcal I}_{d+k,d}  (G,B,Y;s,\kappa)
={\rm R.N.}\, \int_{{\mathcal F}} \de\mu\, \varGamma_{d+k,d} (G,B,Y;\tau)
\, {\mathcal F} (s, \kappa , -\tfrac{k}{2};\tau)\ ,
\label{mainint}
\ee
where the Siegel-Narain theta series is given by \cite{Narain:1985jj}
\be
 \varGamma_{d+k,d} (G,B,Y;\tau) = \tau_2^{\tfrac{d}{2}} \sum_{(m_i,n^i,q^a) \in \Lambda}
 q^{\tfrac14 p_L^2}\, \bar q^{\tfrac14 p_R^2}\ .
\ee
Here, $m_i,n^i$ run over integers while $q^a$ takes values in an even self-dual Euclidean lattice $\Lambda_{\rm E}$ of dimension $k$ (hence $k$
must be a multiple of 8, and $\Lambda=\IZ^{d,d}\oplus \Lambda_{\rm E}$), $p_L^2-p_R^2=4(m_i n^i + \tfrac12(q^a)^2)$ and $p_L^2+p_R^2=\cM^2(m_i,n^i,q^a)$ is a positive definite quadratic form on $\Lambda$ parametrized by the Grassmaniann $G_{d+k,d}$, coordinatized by $(G_{ij}, B_{ij}, Y_i^a)$.  
It is worth noting that the lattice partition function satisfies the differential
equation \cite{Obers:1999um}
\begin{equation}
\label{DelSOG}
\left[ \Delta_{G_{d+k,d}} -2\, \Delta_{\cH,-k/2} + \tfrac14 \,d(d+k-2)-\tfrac{k}{2} \right]\, \varGamma_{d+k,d} = 0\, .
\end{equation} 
where $\Delta_{G_{d+k,d}}$ is the Laplace-Beltrami operator on $G_{d+k,d}$.

Due to the exponential growth near the cusp, the integral must be regulated by 
truncating the fundamental domain to $\cF_\cT$ and taking the limit $\cT\to\infty$,
as in \eqref{renormint}. The integral over $\cF_\cT$ can be computed using the 
generalized unfolding trick \eqref{gentrick}. Carrying out these steps, one finds
that the modular integral \eqref{mainint}, away from the loci in $G_{d+k,d}$ where
one of the lattice vectors becomes null ($p_{\rm R}^2=0$) can be written as an  
infinite sum  \cite{Angelantonj:2012gw,1004.11021}\footnote{For the values $s=1-\tfrac{w}{2}+n$ relevant for the expansion
\eqref{Phiexp}, the summand in \eqref{int2F1} can be rewritten in terms of elementary 
functions \cite{Angelantonj:2012gw}.}
\be
\begin{split}
{\mathcal I}_{d+k,d} (s,\kappa ) 
=&  \sum_{\rm BPS} \, 
 \int_0^{\mathcal \infty} \frac{\de\tau_2}{\tau_2^2} \, \tau_2^{d/2}\, 
 {\mathcal M}_{s,-\frac{k}{2}} (-\kappa \tau_2 )\, 
e^{-\pi\tau_2 (p_{\rm L}^2 + p_{\rm R}^2)/2}\, \\
=&
(4\pi \kappa )^{1-\frac{d}{2}}\, \varGamma (s+ \tfrac{2d+k}{4} -1) 
\\
&\times \sum_{\rm BPS}  \,\, {}_2 F_1  \left(s-\tfrac{k}{4} \,,\, s+ \tfrac{2d+k}{4} -1 \,;\, 2s \,;\, \tfrac{4 \kappa}{ p_{\rm L}^2} \right)\, \left(\frac{p_{\rm L}^2}{4\kappa} \right)^{1-s- \frac{2d+k}{4}}  
\end{split}
\label{int2F1}
\ee
where the sum runs over $(m_i,n^i,q^a)\in \Lambda$ subject to the quadratic `BPS' constraint
\be
\label{BPScons}
p_L^2 - p_R^2 = 4(m_i n^i + \tfrac12 (q^a)^2) = 4\kappa \ .
\ee
The unfolding method shows that the sum converges  absolutely for $\Re(s)>\frac{2d+k}{4}$  (away
from afore-mentioned loci) and has a meromorphic  continuation to $\Re(s)>1$,  
 with a simple pole at $s=\frac{2d+k}{4}$ \cite{1004.11021}. Thus ${\mathcal I}_{d+k,d} (s,\kappa )$ defines a function on the Grassmannian $G_{d+k,d}$, manifestly invariant under 
 the automorphism group $SO(d+k,d,\IZ)$ of the lattice $\Lambda$, and  eigenmode
of $ \Delta_{G_{d+k,d}}+$, as a consequence of 
 \eqref{DelSOG} and \eqref{lapF}, 
 \begin{equation}
\label{LapsEpstein}
\begin{split}
\left[ \Delta_{G_{d+k,d}}+\tfrac{1}{16}\, (2d+k-4s)(2d+k+4s-4) \right]\, 
{\mathcal I}_{d+k,d} (s,\kappa )  =0\ .
\end{split}
\end{equation}
At the value $s=\frac{2d+k}{4}$, the eigenvalue vanishes but the renormalized integral \eqref{mainint} must be defined by subtracting the pole. The finite reminder $\hat{\mathcal I}_{d+k,d} (\frac{2d+k}{4},\kappa )$ is then a quasi-harmonic function on $G_{d+k,d}$, mapped to a constant function
by the Laplacian $\Delta_{G_{d+k,d}}$. 

Physically, \eqref{int2F1} is interpreted as a sum of field-theoretical one-loop amplitudes, 
with BPS particles of mass $p_L^2+p_R^2-4\kappa$ propagating in the loop. The singularities
on the loci where a lattice vector becomes null, $p_R^2=0$ originate from one of these particles becomes massless.  The manifestly T-duality invariant BPS sum \eqref{LapsEpstein} should be contrasted from the result obtained in  \cite{Harvey:1995fq,Harvey:1996gc} by the lattice unfolding method. It is worth noting that the result \eqref{int2F1} generalizes easily to modular integrals
of Niebur-Poincar\'e series times lattice partition functions with momentum insertions \cite[\S3.3]{Angelantonj:2012gw}.

\subsection{Fourier-Jacobi expansion}

While the BPS state sum \eqref{int2F1} is manifestly invariant under T-duality and exhibits
singularities from massless states in a transparent fashion, it is in general non-trivial to extract
the asymptotic expansion at a particular boundary component of the Narain moduli space
$SO(d+k,d,\IZ)\backslash G_{d+k,d}$, i.e. at infinity in a  particular Weyl chamber. Of course,
this asymptotic expansion is precisely what is provided by the  lattice unfolding method. 
In this section, we shall explain how to extract it from the BPS sum \eqref{int2F1}, in the special
case $d=2, k=0,\kappa=1$ \cite{AFP4-to-appear}. The generalization to $\kappa\neq 1$ is straightforward,
but the extension to asymmetric lattices with $k>0$ will be discussed in  \cite{AFP4-to-appear}.

For two-dimensional lattices with $\kappa=1$, the BPS constraint \eqref{BPScons} implies
that the integer matrix 
\be
\gamma= \begin{pmatrix} m_1 & -m_2 \\ n^2 & n^1 \end{pmatrix} = 
\begin{pmatrix} 1 & \tilde M  \\ 0 & 1 \end{pmatrix} 
\begin{pmatrix} m'_1 & -m'_2 \\ n^2 & n^1 \end{pmatrix} 
\ee
is an element of $SL(2,\IZ)$. As written above, such matrices can be decomposed into
products of an upper triangular matrix with $\tilde M\in \IZ$ and coset representatives 
of $\Gamma_{\infty}\backslash SL(2,\IZ)$. After Poisson resummation over the
integer $\tilde M$, the first line of \eqref{int2F1} can be rewritten as 
\be
\begin{split}
{\mathcal I}_{2,2} (T,U;s,1) =& \sum_{M\in\IZ}\,  
\sum_{\gamma\in \Gamma_{\infty}\backslash \Gamma}\, 
 \int_0^{\mathcal \infty} \frac{\de\tau_2}{\tau_2} \, 
 {\mathcal M}_{s,0} (-\kappa\tau_2 )\,
 \sqrt\frac{T_2 U_2}{\tau_2} \, \\
 &
 \exp\left[-\pi\tau_2 \left( \frac{T_2}{ U_2} + \frac{U_2}{T_2}\right)
 -\frac{\pi M^2 T_2 U_2}{\tau_2} 
 +2\pi\I M (T_1- U_1)\right]\vert_{\gamma}
\end{split}
\ee
where the slash operator $\vert_\gamma$ now acts by replacing $U\mapsto  \frac{m_1' U-m_2'}{n^2 U+n^1}$. Thus, the right-hand side is a sum of Poincar\'e series in $U$, with $T$-dependent coefficients. Evaluating the integral over $\tau_2$ in the chamber where $T_2$ is larger than all $U_2\vert_{\gamma}$,we find
\be
\begin{split}
{\mathcal I}_{2,2} (T,U;s,1) =& 
2^{2s} \, \sqrt{4\pi} \, \Gamma(s-\tfrac12) \, T_2^{1-s} \,  \frac{E^\star(s;U)}{\pi^{-s} \Gamma(s)}
\\
&+4 \sum_{M>0}\, 
\sqrt{\frac{ T_2}{M}} \, K_{s-\tfrac12}(2\pi M T_2)\, \left[ e^{2\pi\I M T_1}
\, \cF(s,M,0;U)  + {\rm c.c} \right]\ .
\end{split}
\label{genTexp}
\ee
This provides the asymptotic expansion of ${\mathcal I}_{2,2} (T,U;s,1)$ near the dimension-one
boundary component $T\to \I\infty$ keeping $U$ fixed and arbitrary. 

For $s=1$, based on \eqref{LapsEpstein} we expect $\hat{\mathcal I}_{2,2} (T,U;1,1)$ to be  a
quasi-harmonic modular form in $(T,U)$. Indeed, one may use 
$K_{1/2}(x) =\sqrt\frac{\pi}{2x} e^{-x}$, \eqref{Fex}, \eqref{DwF} and \eqref{kron1} to obtain
\begin{equation}
\begin{split}
{\mathcal I}_{2,2}(T,U;1,1)
=& -24 \log(4\pi T_2 U_2 |\eta(U)|^4 |\eta(T)|^4) \\
& -8\pi T_2 \ + 2\sum_{N> 0}\, \frac{1}{N}\left[ q_T^N \, H_{N} \cdot J(U) + {\rm c.c.}
 \right]
\label{mainint3}
\end{split}
\end{equation}
The second line is recognized as the real part of the logarithm of Borcherds' infinite 
product \cite[Eq. 7.1]{zbMATH00220742}
\be
\label{BorcherdsTU}
 \log\left[  q_T\, (J(T) - J(U) ) \right] = - \sum_{N>0} \frac{1}{N}\, q_T^N\, H^{(U)}_N \cdot J(U)  \ .
\ee
Combining \eqref{mainint3} and \eqref{BorcherdsTU}, we arrive at  the well-known
 result \cite{Harvey:1995fq} (up to an additive constant)
\begin{equation}
\label{hm1}
{\rm R.N.}\, \int_{{\mathcal F}} \de\mu\, \varGamma_{2,2} (T,U)\, (J(\tau)+24) = - \log | J(T) - J(U) |^4  - 24 \, \log \left[ T_2 U_2 |\eta (T) \, \eta (U) |^4 \right]\ .
\end{equation}

For $s=n+1$ with $n$ integer, one can similarly use the properties
\be
\label{cFdern}
\begin{split}
&2\, (-2N)^n\, \sqrt{N T_2}\, K_{n+\tfrac12}(2\pi N T_2)\, e^{2\pi\I N T_1}
=D_T^n \, q_T^{N} 
\\
&(2\kappa)^n n! \, \cF(n+1,\kappa,0;U)   = 
D_U^n \cF(n+1,\kappa,-2n;U)
\\
&\pi^{n+1} \, E^\star(n+1;U) = (2\pi)^n \, 
D_U^n\, E(n+1,0,-2n;U) \ ,
\end{split}
\ee
to express ${\mathcal I}_{2,2} (n+1,1)$ as 
\be
\label{iderpot}
{\mathcal I}_{2,2} (n+1,1) = 4 \, \Re\left[  \frac{(-D_T  D_U)^n}{n!} f_n(T,U) \right]\ ,
\ee
where $f_n(T,U)$ is a generalized prepotential, which is a linear combination harmonic Maass form of weight $-2n$ in $U$, with coefficients which are holomorphic in $T$,
\be
f_n(T,U) = 2\, (2\pi)^{2n+1}\, E(n+1,0,-2n;U) + 
\sum_{M>0} \frac{2}{(2M)^{2n+1}} q_T^{M}\, H_M^{(U)}\cdot \cF(n+1,1,-2n;U)\ .
\ee
Holomorphicity in $U$ may be restored by replacing $E(n+1,0,-2n;U)$ and $ \cF(n+1,1,-2n;U)$
by their analytic parts, proportional to the Eichler integrals of the usual holomorphic 
Eisenstein series $E_{2n+2}(U)$ and Poincar\'e series $\cF(n+1,1,2n+2;U)$. The resulting
generalized non-holomorphic prepotential $\tilde f_n(T,U)$ will no longer be covariant
under T-duality, but rather transform as an Eichler integral, picking up additional polynomials
of degree $2n$ in $(T,U)$ under $SL(2,\IZ)_T\times SL(2,\IZ)_U\ltimes \sigma_{T,U}$.
For $n=1$, $f_1(T,U)$ describes the one-loop correction to the prepotential in $\cN=2$
heterotic string vacua, and was indeed observed to transform by period integrals in the
prescient paper  \cite{Antoniadis:1995ct}. Generalized prepotentials with $n=2$ also
arose in the study of $F^4$ corrections in $D=8$ heterotic string 
vacua  \cite{Lerche:1998nx,Lerche:1998gz},
and were introduced for general $n$ in \cite{Kiritsis:1997hf,Lerche:1999hg}. Our approach gives a straightforward
derivation of their modular properties.

%%%%%%%%%%%%%%%%%%

\section{Higher-loop modular integrals \label{sec_highergen}}

In this last section, we tackle the case of higher-loop modular integrals of the form
\eqref{modinth2}, which was one of our main motivations for developing the Rankin-Selberg technique. Unfortunately, Siegel-Poincar\'e series of degree $h\geq 2$ are {\it terra incognita}
in the mathematical literature, and we shall content ourselves with modular integrals 
of a symmetric lattice partition function and trivial elliptic genus,
 \be
\label{modinth3}
\cA_h(G,B)= {\rm R.N.} \int_{\cF_h}\, \de\mu_h\, \Gamma_{d,d,h}(G,B;\Omega)\ .
\ee
Our aim will to compute \eqref{modinth3} using the same strategy as in  \S\ref{sec_1looptriv}, by inserting a non-holomorphic Eisenstein series
in the integral, applying the unfolding trick and extracting a suitable residue. 

\subsection{Non-holomorphic Eisenstein series}

Recall that the Siegel upper half plane of degree $h$,
\be
\cH_h = \{\Omega=\Omega_1+\I \Omega_2 \in \IC^{h\times h},\, \Omega=\Omega^t,\,\Omega_2>0\}
\ee
admits a transitive action  $\Omega\mapsto (A\Omega+B)(C\Omega+D)^{-1}$ of the Siegel modular group 
\be
\Gamma_h = Sp(h,\IZ) = \left\{ \gamma=\begin{pmatrix} A & B \\ C & D \end{pmatrix}\in \IZ^{2h\times 2h}\ ,\,
\begin{array}{c} AB^t=BA^t, CD^t=DC^t \\ AD^t-BC^t=1_{h} \end{array}
\right\}\ .
\ee
The completed non-holomorphic Eisenstein series of weight 0 under $\Gamma_h$ is defined by \cite{0397.10021,0786.11024,0715.11025}\footnote{For $h=1$, $E_1^*(s;\Omega)$ reduces
to the Eisenstein series \eqref{Edefinition} for $SL(2,\IZ)=Sp(1,\IZ)$.}
\be
\label{defEh}
E_h^*(s;\Omega) = \cN_h(s)\!\!  \sum_{\gamma\in \Gamma_{h,\infty}\backslash \Gamma_h} |\Omega_2|^s 
\vert_0\gamma\ ,\quad \cN_h(s)=\zetastar(2s) \, \prod_{j=1}^{\lfloor h/2\rfloor} \zetastar(4s-2j)\, 
\ee
where $|\Omega_2|=\det \Omega_2$ and $\vert_w\gamma$ denotes the Petersson slash operator
\be
F\vert_w \gamma (\Omega) = [\det(C\Omega+D)]^{-w}\, F[(A\Omega+B)(C\Omega+D)^{-1}] 
\ee
and $\Gamma_{h,\infty}$ is the subgroup of $\Gamma_h$ of matrices with $C=0$.
Equivalently,
\be
E_h^\star(s;\Omega) =  \cN_h(s)\,  
\sum_{(C,D)\in GL(h,\IZ) \backslash \IZ^{(h,2h) }\atop (C,D)=1} 
\left[ \frac{|\Omega_2|}{|C\Omega+D|^2}\right]^{s}
\ee
where the sum runs over pairs of coprime symmetric integer matrices $(C,D)$, modulo a common left multiplication
by $GL(h,\IZ)$. The sum converges absolutely for $\Re(s)>\tfrac{h+1}{2}$, has a meromorphic
continuation to the $s$-plane, and is  an eigenmode of the Laplace-Beltrami operator on $\cH_h$,
\be
\label{DelEh}
\Delta_{\cH_h}\, E^\star_h(s;\Omega) = \frac12 h s (2s-h-1) \, E^\star_h(s;\Omega)\ .
\ee
With the choice of normalization in \eqref{defEh}, $E^*_h(s;\Omega)$ is invariant under $s\mapsto \frac{h+1}{2}-s$, with poles at most at $s=j/4$ with $0\leq j\leq 2h+2$ \cite{0786.11024}.

The Fourier expansion with respect to $\Gamma_{\infty}$ takes the form
\be
E_h^\star(s;\Omega) =\sum_{\substack{2T\in \IZ^{(h,h)}\\ T_{ii}\in \IZ}} E_h^\star(T;s;\Omega_2)\, e^{2\pi\I\Tr[T\Omega_1]}
\ee
where the sum runs over half-integer symmetric $h\times h$ matrices $T$ (i.e. such that $2T$ is integer with even diagonal entries). The zero-th Fourier mode  is given by \cite{0786.11024}
\be
\label{Estarh00}
\begin{split}
E_h^\star(T=0;s;\Omega_2) = &
\sum_{r=0}^h
\zetastar(2s-r) \prod_{j=\lceil h/2\rceil}^{r-1} \zetastar(4s-2j-1)\, \prod_{j=r+1}^{\lfloor h/2\rfloor} {\zetastar(4s-2j)}\,\\
& \times |\Omega_2|^{s-\tfrac{r}{h}(2s-\tfrac{r+1}{2})}\,
\cE^{\star;SL(h,\IZ)}_{\Lambda^r V}\left(  2s-\tfrac{r+1}{2}; \hat\Omega_2\right) 
\end{split}
\ee
where $\hat \Omega_2=\Omega_2/|\Omega_2|^{1/h}$ and $\cE^{\star;SL(h,\IZ)}_{\Lambda^r V}$ is the completed Langlands-Eisenstein series
of $SL(h,\IZ)$ with infinitesimal character $\rho-2s\lambda_r$, where $\lambda_r$ is
the weight associated to the $r$-fold antisymmetric product of the defining representation,
\be
\label{Estarh0}
\cE^{\star;SL(h,\IZ)}_{\Lambda^r V}\left( s;\hat g \right) = 
\prod_{j=0}^{r-1} \zetastar(2s-j)\, \sum_{Q\in\IZ^{h\times r}_{\rm prim}/GL(r,\IZ)} 
[\det( Q^t \hat g Q)]^{-s}\ ,
\ee
with the understanding 
that $\cE^{\star;SL(h,\IZ)}_{\Lambda^0 V}=\cE^{\star;SL(h,\IZ)}_{\Lambda^h V}=1$. 
In \eqref{Estarh0} the sum runs over primitive integer $h\times r$ matrices $Q$
modulo right action of $GL(r,\IZ)$. It  
satisfies the functional equation
\be
\cE^{\star;SL(h,\IZ)}_{\Lambda^r V}\left( s;\hat g \right) =
\cE^{\star;SL(h,\IZ)}_{\Lambda^{h-r} V}\left( \tfrac{h}{2}-s ; \hat g\right) \ .
\ee
Most importantly, $E_h^\star(s;\Omega)$ has a  simple pole with constant residue $r_h$ 
at $s=\frac{h+1}{2}$ (and consequently a simple pole with at $s=0$ with residue $-r_h$) 
where
\be
\label{rescst}
r_h = - {\rm Res}_{s=0} \, \cN_h(s) = \frac12 \, \prod_{j=1}^{\lfloor h/2\rfloor}\, \zetastar(2j+1)\ ,
\ee
which can be read off from the terms with $r=0$  in \eqref{Estarh00}.

\subsection{Rankin-Selberg method}

For a non-holomorphic modular form $F(\Omega)$ of weight 0 and of rapid decay at the cusp, the 
modular integral
\be
\cR^\star_h(F;s) = \int_{\cF_h}\,\de\mu_h \, E^\star_h(s;\Omega)\, F(\Omega) 
\ee
over a fundamental domain $\cF_h$ of the Siegel upper half plane is convergent whenever
$\Re(s)>h+1$, and can be computed by the  unfolding trick:  the sum over 
$\Gamma_{\infty}\backslash \Gamma$ is traded for an integral over the `generalized strip'
\be
\label{defSh}
\cS_h=\Gamma_{\infty}\backslash \Gamma_h=
GL(h,\IZ)\backslash(\cP_h\times [-\tfrac12,\tfrac12]^{h(h+1)/2})\ ,
\ee
where $\cP_h=GL(h,\IR)/SO(h)=\IR^+\times SL(h,\IR)/SO(h)$ is the
space of positive definite symmetric real matrices. Integrating  along $\Omega_1$
replaces $F(\Omega)$ by its zeroth Fourier coefficient $F_0(\Omega_2)=\int_{0}^1 \de\Omega_1 F(\Omega)$, leading to 
\be
\label{RStransform}
\cR^\star_h(F;s) =  \cN_h(s)\, 
\int_{GL(h,\IZ)\backslash\cP_h} \frac{\de\Omega_2}{|\Omega_2|^{h+1-s}} F_0(\Omega_2)\ .
\ee
The integration domain $GL(h,\IZ)\backslash\cP_h$ is the product of a semi-infinite line
$\IR^+$, associated to the determinant $|\Omega_2|$, times a fundamental domain for the action 
of $SL(h,\IZ)$ on the space of positive definite symmetric real matrices of determinant one, e.g.
the one constructed by Minkowski \cite{zbMATH02628371}. 

The Rankin-Selberg transform, defined by \eqref{RStransform}, inherits the analytic properties of $ E^\star_h(s;\Omega)$, in particular
it is meromorphic in $s$ with a simple pole at $s=0,\tfrac{h+1}{2}$ and satisfies the functional equation
\be
\cR^\star_h(F;s) = \cR^\star_h(F;\tfrac{h+1}{2} - s) \ .
\ee
Since the residue of $E^\star_h(s;\Omega)$ is a constant \eqref{rescst}, 
the modular integral of $F$ over $\cF_h$ is proportional to the residue of $\cR^\star_h(F;s)$ at the same point,
\be
 \int_{\cF_h} \, \de\mu\, F = \frac{1}{r_h} {\rm Res}_{s=\tfrac{h+1}{2}} \cR^\star_h(F;s)\ .
\ee

\subsection{Higher-loop BPS state sums}

The Rankin-Selberg method described in the previous subsection is, unfortunately,
not directly applicable to the modular integral \eqref{modinth3}, since the Siegel-Narain theta series \eqref{defSiegel} is not of rapid decay at $\Omega_2\to\infty$. This is best seen after expliciting \eqref{defSiegel} as
\be
\label{Gamddh}
\Gamma_{d,d,h}(G,B;\Omega) =|\Omega_2|^{d/2} \!\!\!\!\!
 \sum_{(m_i^\alpha,n^{i\alpha})\in\IZ^{2d}}
   \!\!\!\!\!
 e^{-\pi \Tr(\cM^2 \Omega_2)+2\pi\I m_i^{\alpha} n^{i\beta} \Omega_{1,\alpha\beta}}
\ee
where
\be
\cM^{2;\alpha\beta}=(m_i^\alpha+B_{ik} n^{k\alpha}) G^{ij}  (m_j^\beta+B_{jl} n^{l\beta}) + n^{i\alpha} G_{ij}  n^{j\beta} 
\ee
is the Gram matrix of the positive definite quadratic form \eqref{M2GB} on $h$-tuples of vectors $(m_i,n^i)^{\alpha}$ in $\IZ^{2d}$. The $h$-tuple contributes to the zero-th Fourier coefficient  
$F_0(\Omega_2)$ of $\Gamma_{d,d,h}$ whenever  $m_i^{(\alpha} n^{i\beta)}=0$ for 
all $\alpha,\beta$, i.e. when the $h$ vectors  $(m_i,n^i)^{\alpha}$ span an isotropic subspace
of $\IR^{d,d}$. The contribution is exponentially suppressed as $\Omega_2\to\infty$ unless 
 the Gram matrix $\cM^{2;\alpha\beta}$ has vanishing determinant, i.e. when the $h$ vectors 
 $(m_i,n^i)^{\alpha}$ are linearly dependent. Since the dimension of the maximal isotropic subspace of $\IR^{d,d}$ is $d$, this is always the case if $d<h$. As in the genus one case 
 \eqref{defRstar}, it is natural to extend the definition of the Rankin-Selberg transform \eqref{RStransform} by subtracting the non-decaying part of $F_0(\Omega_2)$, leading to
 \be
\label{RStransform2}
\cR_h^\star(\Gamma_{d,d,h};s) = \cN_h(s)\, 
\int_{GL(h,\IZ)\backslash\cP_h} \frac{\de\Omega_2}{|\Omega_2|^{h+1-s-\tfrac{d}{2}}} 
\!\!
 \sum_{(m_i^\alpha,n^{i\alpha})\in\IZ^{2d\times h} \atop  m_i^{(\alpha} n^{i\beta)}=0, 
 {\rm Rk}(m_i^{\alpha},n^{i\beta})\geq h}
   \!\!\!\!\!
 e^{-\pi \Tr(\cM^2 \Omega_2)}\ ,
 \ee
where the  sum is empty if $d<h$.
By the unfolding trick again, this can be written as an integral over the full space of
positive definite symmetric matrices $\cP_h$, at the expense of restricting the sum
to $GL(h,\IZ)$ orbits,  
 \be
 \label{RStransform3}
 \cR_h^\star(\Gamma_{d,d,h};s) = \cN_h(s)\, 
\int_{\cP_h} \frac{\de\Omega_2}{|\Omega_2|^{h+1-s-\tfrac{d}{2}}} 
\!\!
 \sum_{(m_i^\alpha,n^{i\alpha})\in\IZ^{2d\times h}/GL(h,\IZ) \atop  m_i^{(\alpha} n^{i\beta)}=0, 
 {\rm Rk}(m_i^{\alpha},n^{i\beta})\geq h}
   \!\!\!\!\!
 e^{-\pi \Tr(\cM^2 \Omega_2)}\ .
\ee
This integral can be carried out using  \cite{0066.32002} 
\be
\label{GenEuler}
\int_{\cP_h} \, \de \Omega_2\, | \Omega_2|^{\delta-\frac{h+1}{2}} \, 
e^{-\Tr( Q \Omega_2)} = \Gamma_h(\delta)\, | Q|^ {-\delta}\, \quad
\ee
where  the right-hand side is fixed by invariance under $SL(h,\IR)$ and dimensional analysis, up to a multiplicative factor given by 
 \be
 \Gamma_h(s) = \prod_{k=0}^{h-1} \pi^{k/2} \Gamma(s-\tfrac{k}{2}) \ .
 \ee
Using \eqref{GenEuler}, we find that the regularized  Rankin-Selberg transform of 
$\Gamma_{d,d,h}$ is given by the `higher genus BPS sum'
\be
\label{Rhdet}
\cR_h^\star(\Gamma_{d,d,h};s)  =  \cN_h(s)\, \Gamma_h\left(s-\tfrac{h+1-d}{2}\right)\, 
\sum_{\rm BPS}\, 
\left[ \det( \cM^2 ) \right]^{-s+\frac{h+1-d}{2}}
\ee
where
\be
\label{sumBPS}
\sum_{\rm BPS} =  \sum_{(m_i^\alpha,n^{i\alpha})\in\IZ^{2d\times h}/GL(h,\IZ) 
\atop  m_i^{(\alpha} n^{i\beta)}=0, 
 {\rm Rk}(m_i^{\alpha},n^{i\beta})\geq h}\ .
\ee
For $d>h$, this is recognized as the degenerate Langlands-Eisenstein series 
of $SO(d,d,\IZ)$ with infinitesimal parameter $\rho-2s' \lambda_h$,
with $s'=s-\frac{h+1-d}{2}$,  attached to the representation $\Lambda^h V$ where $V$ is the vector representation,
\be
\label{eqRhSO}
\cR_h^\star(\Gamma_{d,d,h};s) = \cE^{SO(d,d),\star}_{\Lambda^h V}(s-\tfrac{h+1-d}{2};G,B) \ .
\ee
For $d=h$,  the representation  $\Lambda^h V$ decomposes into a sum of two irreps
with weight $2\lambda_S$ and $2\lambda_C$ where $\lambda_S,\lambda_C$ are the
weights associated to the two inequivalent spinor representations, and \eqref{eqRhSO}
continues to hold if we define
\be
\label{Rhd}
\cE^{SO(h,h),\star}_{\Lambda^h V}(s)=
\prod_{k=0}^h \zetastar(2s+1-k)\, \left[ \cE^{SO(h,h),\star}_{S}(2s) + \cE^{SO(h,h),\star}_{C}(2s)\right]\ . 
\ee
These identifications are consistent with the fact that $\cR_h^\star(\Gamma_{d,d,h};s)$ is an
eigenmode of the Laplace-Beltrami operator on $G_{d,d}$ with eigenvalue
\be
\left[ \Delta_{G_{d,d}} -\tfrac{h}{4}(2s-d)(2s+d-h-1) \right]\, \cR^\star_h(\Gamma_{d,d,h};s) = 0\ ,
\ee
as follows from  \eqref{DelEh} and the generalization of \eqref{DelSOG} 
to genus $h$ \cite{Obers:1999um},
\be
\left[\Delta_{G_{d,d}} - \Delta_{\cH_h} + \tfrac{d h(d-h-1)}{4} \right]\, \Gamma_{d,d,h} = 0\ .
\ee
The functional equation $E^\star_h(s)=E^\star_h(\tfrac{h+1}{2}-s)$ implies that
$\cE^{SO(d,d),\star}_{\Lambda^h V}(s)$ is invariant under $s\mapsto d-\tfrac{h+1}{2}-s$.

\subsection{Higher loop string and field theory amplitudes}

By a similar reasoning as in \S\ref{sec_rsz}, the modular integral  \eqref{modinth3} should
be proportional to the residue of the regularized Rankin-Selberg transform $\cR^\star_h(\Gamma_{d,d,h};s)$, up to a renormalization scheme-dependent subtraction $\delta$,
\be
\label{Resdel}
{\rm Res}_{s=\tfrac{h+1}{2}} \cR^\star_h(\Gamma_{d,d,h};s)
=\frac{1}{r_h} {\rm R.N.} \int_{\cF_h} \de\mu\, \Gamma_{d,d,h} + \delta\ .
\ee
Unfortunately, we have not yet been able to imitate the method of \cite{MR656029}
to  compute the subtraction $\delta$. Using invariance under $s\mapsto \tfrac{h+1}{2}-s$
and assuming that the simple pole at $s=0$ arises entirely from the prefactor $\cN_h(s)$,
we obtain
\be
\label{AhRS}
\cA_h= \delta + 
\int_{GL(h,\IZ)\backslash\cP_h} \frac{\de\Omega_2}{|\Omega_2|^{h+1-\tfrac{d}{2}}} 
\!\!
 \sum_{(m_i^\alpha,n^{i\alpha})\in\IZ^{2d\times h} \atop  m_i^{(\alpha} n^{i\beta)}=0, 
 {\rm Rk}(m_i^{\alpha},n^{i\beta})\geq h}
   \!\!\!\!\!
 e^{-\pi \Tr(\cM^2 \Omega_2)}\ .
\ee
It is interesting to observe that the contribution of the terms with zero winding, $n^{i\alpha}$=0,
is exactly of the form expected for a $h$-loop amplitude in ten-dimensional supergravity compactified on a torus $T^d$, with the integration domain $GL(h,\IZ)\backslash\cP_h$ being
identified as the space of Schwinger parameters. For $h=2$, it was indeed noted in the context
of $D^4 \cR^4$ couplings in eleven-dimensional supergravity \cite{Green:1999pu} that the three Schwinger parameters $L_1,L_2,L_3$ could be mapped by a variable change
\be
V=(L_1L_2+L_2L_3+L_1L_3)^{-1/2}\ ,\quad
\tau_1=\frac{L_2}{L_2+L_3}\ ,\quad \tau_2=\frac{1}{V(L_2+L_3)}\ ,\quad 
\ee
to $\IR^+_V \times \cF_{\Gamma_0(2)}$, where $\cF_{\Gamma_0(2)}$ is the fundamental domain
of the action of the Hecke subgroup $\Gamma_0(2)$ on the Poincar\'e upper half plane
parametrized by $\tau$. Using the invariance of the integrand under a larger group $SL(2,\IZ)$,
it was shown that the unregulated two-loop field theory integral can be written as 
\be
\label{AFT2}
\cA_{2}^{\rm F.T.} = \int_{GL(2,\IZ)\backslash\cP_2}  
\frac{\de\Omega_2}{|\Omega_2|^{3-\tfrac{d}{2}}} 
\!\! \sum_{m_i^\alpha\in\IZ^{d\times h}}    \!\!
 e^{-\pi \Tr(m_i^\alpha \Omega_{2,\alpha\beta} m^j_\beta G_{ij})}\
\ee
with 
\be
\label{Om2tau}
\Omega_2= \begin{pmatrix} L_1+L_2 & L_2 \\ L_2 & L_2+L_3 \end{pmatrix}
= \frac{1}{V\tau_2} \begin{pmatrix} |\tau|^2 & \tau_1 \\ \tau_1 & 1 
\end{pmatrix} 
\ee
This indeed matches the zero-winding contribution to \eqref{AhRS}, with the understanding
that $\delta$ incorporates contributions with ${\rm Rk}(m_i^\alpha)<2$, which are responsible
for infrared (and, in field theory, ultraviolet) divergences.  It would be interesting to perform
a similar matching for the $D^6 \cR^4$ couplings in type II on $T^d$ , proportional
to the integral \eqref{modinth3} at 3 loops \cite{Green:2005ba,Gomez:2013sla}\footnote{Note added in v2: Progress on $D^6 \cR^4$ couplings at two and three loops were recently reported in 
\cite{D'Hoker:2014gfa, Basu:2014hsa}.}.

\subsection{Lattice unfolding method}

While we are not able to compute the subtraction $\delta$ yet, in this section and the next
we shall compute the modular
integral using the lattice unfolding method, and comparing with the Rankin-Selberg transform
in cases where both results are available in closed form. For this purpose, we shall use
the Lagrangian representation of the lattice partition function, where modular invariance is manifest, 
\be
\begin{split}
\Gamma_{d,d,h}&(G,B;\Omega) = V_d^h\, \!\!\!\!\sum_{(M,N)\in \IZ^{2hd}} \\
&\!\!\!\!
\exp\left( -\pi G_{ij}(M^i_\alpha-\Omega_{\alpha\beta} N^{i\beta}) [\Omega_2^{-1}]^{\alpha\gamma} 
(M^j_\gamma-\bar\Omega_{\gamma\delta} N^{j\delta})+2\pi\I B_{ij} M^i_\alpha N^{j\alpha}\right)\ .
\end{split}
\ee
where $V_d=\sqrt{\det{G_{ij}}}$.
This expression follows from \eqref{Gamddh} by Poisson resummation on $m_i^\alpha$,
and is manifestly invariant under $Sp(h,\IZ)$ action on $\Omega$, with $(M^i_{\alpha},N^{i\alpha})$ transforming in the defining representation of $Sp(h,\IZ)$ for any $i=1\dots d$.
Orbits under $Sp(h,\IZ)$ are classified (in part) by the rank of the $d\times 2h$ matrix $(M^i_{\alpha},N^{i\alpha})$ and by the $d\times d$ antisymmetric matrix $m^{ij}=M^i_\alpha N^{j\alpha}$.

\subsubsection{Zero orbit}

The term with $M^i_\alpha=N^{i\alpha}=0$ is invariant under the action of $Sp(h,\IZ)$.
 Its integral over $\cF_h$ is proportional to the volume $\cV_h$ of the fundamental domain,
 \be
 \label{Azero}
\cA^{(0)}_h =  \cV_h\, V_d^h \ . 
\ee
where $\cV_h$ is given by \cite{0063.07003}
\be
\label{volh}
\cV_h= \int_{\cF_h}\, \de\mu_h = 2 \, \prod_{j=1}^h\, \zetastar(2j)\ ,
\ee
so that $\cV_1=\pi/3$, $\cV_2=\pi^3/270$, $\cV_h= \zetastar(h)\, \cV_{h-1}$ whenever $h\geq 2$.

\subsubsection{Rank one orbit}
If $(M^i_{\alpha},N^{i\alpha})$ has rank one, it can be mapped by $Sp(h,\IZ)$  to an orbit
representative with $M^i_{\alpha}=0$ unless $\alpha=h$, and $N^{i\alpha}=0$ for all $i,\alpha$.
The stabilizer of such an element is the Fourier-Jacobi subgroup spanned by
\be
\Gamma_J = \{ \begin{pmatrix} A & 0 & B & 0\\
0 & 1 & 0 & 0\\
C & 0 &D & 0\\
0 & 0 & 0 & 1 \end{pmatrix} \cdot
\begin{pmatrix} 1_{h-1} & 0 & 0 & \mu \\
0 & 1 & \mu^t & \kappa \\
0 & 0 &1_{h-1} & 0\\
0 & 0 & 0 & 1 \end{pmatrix} \cdot
\begin{pmatrix} 1_{h-1} & 0 & 0 & 0 \\
\lambda^t & 1 & 0 & 0 \\
0 & 0 &1_{h-1} &-\lambda \\
0 & 0 & 0 & 1 \end{pmatrix}\}
\ee
where ${\scriptsize \begin{pmatrix} A & B \\ C & D  \end{pmatrix}}$ runs over elements of 
$Sp(h-1,\IZ)$, $\lambda,\mu\in \IZ^{h-1}, \kappa\in \IZ$. Decomposing
the period matrix 
\be
\label{omhm1}
\Omega = \begin{pmatrix} \rho_1 & \rho_1 u_2 - u_1 \\u_2^t \rho_1- u_1^t & \sigma_1\end{pmatrix}
+ \I \begin{pmatrix} \rho_2 & \rho_2 u_2  \\u_2^t \rho_2 & t+ u_2^t \rho_2 u_2 \end{pmatrix}\ ,
\ee
where $t\in \IR^+, \rho=\rho_1+\I\rho_2\in \cH_{h-1}$, $u_1,u_2\in \IR^{h-1}, \sigma_1\in \IR$, the
measure on $\cH_h$ can be written as
\be
\frac{\de\Omega\de\bar\Omega}{|\Omega_2|^{h+1}}= \frac{\de\rho \de\bar\rho}{|\rho_2|^h}\, \frac{\de t}{t^{h+1}}\, \de u_1 \de u_2 \de \sigma_1\ .
\ee
At the cost of restricting to an orbit representative of the above form, the integration domain $\cF_h$ can therefore be unfolded
unto $\Gamma_J\backslash \cH_h = \IR^+ \times \cF_{h-1} \times T^{2h-1}$,
where $T^{2h-1}$ is a twisted torus parametrized by $u_1,u_2,\sigma_1$.
Denoting $M^i_{h}=m^i$, we find
\be
\label{Aone}
\begin{split}
\cA^{(1)}_h =&   \frac12\sum_{m^i\neq 0} V_d^h \, \int_{\cF_{h-1}} \de\mu_{h-1}
\, \int_{T^{2h-1}} \de u_1 \de u_2 \de\sigma_1 \, 
 \int_0^{\infty} \frac{\de t}{t^{h+1}}
 e^{-\pi m^i g_{ij} m^j/t}\\
=&\frac12 \cV_{h-1}\, V_d^h \, \pi^{-h}\, \Gamma(h)\,  \sum_{m^i\neq 0} [m^i G_{ij} m^j]^{-h}\\
=& \frac12 \cV_{h-1}\, V_d^{h(1-\tfrac{2}{d})}\, \cE^{\star;SL(d)}_{V,s=h}(\hat G)
\end{split}
\ee
where $\cE^{\star;SL(d)}_{V,s}(\hat G)$ is the completed Epstein zeta series in the vector representation, evaluated at $\hat G=G_{ij}/[\det(G)]^{2/d}$.

\subsubsection{Rank $h$ orbit with $N^i_\alpha=0$}

If $N^i_\alpha=0$ and $M^{i\alpha}$ is a generic matrix of rank $h$, then the stabilizer
of $(M^{i\alpha},0)$ is the subgroup of matrices with $A=D=1, C=0$. The integral
can be unfolded unto the generalized strip \eqref{defSh}, and after a trivial integration over
$\Omega_1$, produces 
\be
\begin{split}
\label{Ah0}
\cA^{(h)}_h =&  V_d^h   \int_{GL(h,\IZ)\backslash \cP_{h}}\frac{\de\Omega_2}
{|\Omega_2|^{h+1}}
\sum_{  M^i_\alpha\in \IZ^{h\times d} \atop {\rm Rk}(M^i_\alpha)=h}
\exp\left( - \pi G_{ij} M^i_\alpha [\Omega_2^{-1}]^{\alpha\beta} M^i_\beta  \right) \\
=&  V_d^h   \int_{\cP_{h}}  \frac{\de\Omega_2}{|\Omega_2|^{h+1}}
\sum_{ M^i_\alpha\in \IZ^{h\times d}/GL(h,\IZ) \atop
{\rm Rk}(M^i_\alpha)=h} 
\exp\left( - \pi G_{ij} M^i_\alpha [\Omega_2^{-1}]^{\alpha\beta} M^i_\beta  \right) 
\end{split}
\ee
Ignoring for a moment the constraint ${\rm Rk}(M^i_\alpha)=h$ and  performing a Poisson resummation $M^i_\alpha\mapsto m_i^\alpha$, we observe that the first line matches (for genus 2, and presumably genus 3 as well) the field theory 
amplitude \eqref{AFT2}. The integral over $\cP_{h}$ can be computed using \eqref{GenEuler},
and yields
\be
=V_d^h \, \Gamma_h(\tfrac{h+1}{2}) \sum_{ M^i_\alpha\in \IZ^{h\times d}/GL(h,\IZ) \atop {\rm Rk}(M^i_\alpha)=h} 
\, \left[ \det M^i_\alpha G_{ij} M^i_\beta \right]^{-\tfrac{h+1}{2}}
\ee
For $d=h$, the sum over $M^i_\alpha$ can be further evaluated using  \cite{0013.24901}
\be
\label{detsum}
 \sum_{M\in M_h(\IZ)/GL(h,\IZ)} |M|^{-s} = 
 \zeta(s) \zeta(s-1) \dots \zeta(s-h+1)
\ee
Defining $\zeta^\star_h(s) \equiv \prod_{k=0}^{h-1} \zetastar(s-k)$, we find
\be
\cA^{(h)}_h  = V_d^{-1} \zetastar_h(h+1)\ .
\ee

\subsubsection{Orbits with $m^{ij}\neq 0$}

The orbits above all had $m^{ij}=M^i_\alpha N^{j\alpha}$, hence led to contributions independent
of the B-field $B_{ij}$. For $d\geq 2h$, the generic orbit with $m^{ij}\neq 0$ breaks $Sp(h,\IZ)$ entirely, hence can be unfolded on the full Siegel upper-half plane $\cH_h$. The integrals
over $\Omega_{1\alpha\beta}$ are Gaussian, while the integral over $\Omega_{2\alpha\beta}$
can be expressed in terms of the matrix Bessel function of \cite{0066.32002}. There are
also contributions from orbits which leave part of $Sp(h,\IZ)$ unbroken. We shall
not attempt to classify these orbits in full generality, instead we focus on some simple
cases where the full integral is within reach.  

\subsection{Some simple cases}

\subsubsection{$d=1$, any $h$}

For $d=1$, the rank 0 and 1 are the only possible orbits. Using \eqref{Azero} and \eqref{Aone}
we arrive at
\be
\label{Ah1}
\cA_h^{d=1} =\cV_h\, R^{h} + \zetastar(h)\, \cV_{h-1} R^{-h}=  \cV_h\, (R^{h} + R^{-h})\ ,
\ee
in accordance with T-duality. Conversely, T-duality can be used to prove
the recursion formula $\cV_h= \zetastar(h)\, \cV_{h-1}$ hence  \eqref{volh}.
For $h=1$ the same result follows from the Rankin-Selberg transform \cite{Angelantonj:2011br} 
\be
\cR^\star_1(\Gamma_{1,1,1};s)=2\zetastar(2s)\zetastar(2s-1) (R^{1-2s}+R^{2s-1})\ .
\ee
The Rankin-Selberg transform vanishes for $h>1$, and \eqref{Ah1} should originate  
entirely from the subtraction $\delta$ in \eqref{Resdel}. 

\subsubsection{$d=h=2$}

We now consider the genus-two amplitude on $T^2$. By an $Sp(2,\IZ)$ rotation
one can choose 
\be
(M^{i}_\alpha,N^{i\alpha})=\begin{pmatrix} 
0 & p & 0 & 0\\
j_1 & j_2 & j_3 & q \end{pmatrix}\ .
\ee
If $p\neq 0$, the choice of the first vector $(M^{1}_\alpha,N^{1\alpha})$ breaks
$Sp(2,\IZ)$ to the Fourier-Jacobi group $\Gamma_J=SL(2,\IZ)\ltimes \IZ^{2}\ltimes \IZ$.
If $q=0$ one can set $j_3=0$ by means of an $SL(2,\IZ)$ transformation. If $j_1=0$,
$(M^{i}_\alpha,N^{i\alpha})$ has rank 1 case so \eqref{Aone} applies. If $j_1\neq 0$,
$M^i_\alpha$ has rank 2 and $N^{i\alpha}$ vanishes so \eqref{Ah0} applies instead.
Including the  zero orbit, we find that the contributions with $m^{ij}=pq=0$ sum up to
\be
\cA_2^{\rm deg} = \zetastar(2)\zetastar(4)\, T_2^2 +  \zetastar(2)\, E_1^\star(2;U) + 
\zetastar(2) \zetastar(3) T_2^{-1} \ .
\ee
If $pq\neq 0$, such that we can choose $0\leq j_1,j_2,j_3<|q|$ by means of a $ \IZ^{2}\ltimes \IZ$
transformation. Using the parametrization \eqref{omhm1}, the integration domain then unfolds onto $\IR^+(t)\times \cF_1(\rho)\times \IR^3(u_1,u_2,\sigma_1)$. The integral over $u_1,\sigma_1$ and $u_2$ (performed in this order) is Gaussian, with a saddle point at
\be
u_1=\frac{j_1}{q}\ ,\quad u_2=-\frac{j_3}{q}, \quad \sigma_1=
\frac{p U_1}{q|U|^2}+ \frac{j_1j_3-j_2 q+j_3^2\rho_1}{q^2}
\ee
 leading to
\be
\begin{split}
\cA_2^{\rm n.d.} =& \sum_{p,q\atop (p,q)\neq (0,0)} |q|^3 \int_{\cF_1} \frac{\de\rho \de\bar\rho}{\rho_2^2}\ \int_0^{\infty} \frac{\de t}{t^{3}} 
\frac{\sqrt{tU_2 }}{|q|^3 T2^{3/2} |U|} e^{-\frac{\pi p^2 T_2 U_2}{t|U|^2} -  \frac{\pi q^2 |U|^2 T_2}{U_2 t} +2\pi\I pq T_1}\\
=& 2\zetastar(2) \sum_{p,q} \frac{T_2^{1/2}}{|pq|^{3/2}} |q|^3 \, 
K_{3/2}(2\pi|pq|T_2) e^{2\pi\I p q T_1}
\end{split} 
\ee
where the factor $|q|^3$ in front counts the number of $j_i$'s such that $0\leq j_1,j_2,j_3<|q|$.
In total, we find
\be
\label{A22}
\cA_2^{d=2} = \zetastar(2) \left[  E_1^\star (2;T)+E_1^\star (2;U) \right]\ ,
\ee
which we recognize as the sum of spinor and conjugate spinor Epstein series of $SO(2,2)$ with $s=2$, as conjectured in \cite{Obers:1999um}. In the decompactification limit, setting 
 $T=\I R_1R_2, U=R_1/R_2$ and taking $R_2\to\infty$, we see that $\cA_2^{d=2}$
 grows as $R_2^2$ times $\cA_2^{d=1}(R_1)$, as it should.

Alternatively, we can compute the Rankin-Selberg transform \eqref{Rhdet}, and extract
the residue at $s=3/2$. 
Denoting $(m_i^1,n^{i,1})=(m_i,n^i)$ and $(m_i^2,n^{i,2})=(\tilde m_i,\tilde n^i)$,
the quadratic constraints in the BPS sum \eqref{sumBPS} read
\be
m_1 n^1+m_2 n^2 = 0\ ,\quad 
\tilde m_1 \tilde n^1+\tilde m_2 \tilde n^2 = 0\ ,\quad
m_1 \tilde n^1+m_2  \tilde n^2 +  \tilde m_1 n^1+ \tilde m_2 n^2 = 0\ .
\ee
The first two constraints can be solved as in \cite[\S3.2]{Angelantonj:2011br},
\be
\begin{pmatrix} m_1 & m_2 & n^1 & n^ 2 \\
\tilde m_1 & \tilde m_2 & \tilde n^1 & \tilde n^ 2
\end{pmatrix} =
\begin{pmatrix} c k_1 & c k_2 & -d k_2 & d k_1 \\
\tilde c \tilde k_1 & \tilde c \tilde k_2 & -\tilde d \tilde k_2 & \tilde d \tilde k_1
\end{pmatrix} \ .
\ee
The third constraint requires $(c \tilde d-d \tilde c)(k_2 \tilde k_1-k_1 \tilde k_2)=0$,
while the condition ${\rm Rk}(m_i^{\alpha},n^{i\beta})\geq 2$ requires one of the two factors
in this product to be non-vanishing. There are therefore 
two possible branches:
\be
\begin{split}
i) &\quad (\tilde c,\tilde d)= (c,d)\neq (0,0), \quad {\rm gcd}(c,d)=1, \quad  k_2 \tilde k_1-k_1 \tilde k_2 \neq 0 \\
ii) & \quad (\tilde k_1,\tilde k_2) =  (k_1, k_2)\neq (0,0)\ ,
\quad {\rm gcd}(k_1,k_2) = 1\ ,\quad c \tilde d-d \tilde c\neq 0
\end{split}
\ee
In either case,
\be
\begin{split}
i) & \det( \cM^2) = (k_2 \tilde k_1-k_1 \tilde k_2)^2 \frac{|c+dT|^4}{4T_2^2} \\
ii) & \det(\cM^2) = (c \tilde d-d \tilde c)^2 \frac{|k_1+k_2U|^4}{4U_2^2}
\end{split}
\ee
In the first branch, the sum over $c,d$ produces the Eisenstein series $E_1^\star(2s-1;T)/\zetastar(4s-2)$, while the sum over matrices $M={\scriptsize \begin{pmatrix} k_1 & k_2 \\ \tilde k_1 & \tilde k_2 \end{pmatrix}}$ modulo  $GL(2,\IZ)$ can be computed using  \eqref{detsum}.
In total we find
\be
\label{R22}
\cE^{SO(2,2),\star}_{\Lambda^2 V}(s) =  \zetastar(2s+1)\zetastar(2s)\zetastar(2s-1)  
\left[  E_1^\star(2s;T)+E_1^\star(2s;U) \right]\ ,
\ee
in agreement with \eqref{Rhd}.  The residue at $s=1$ reproduces \eqref{A22}, as it should.

\subsubsection{$d=h=3$}
We have not attempted to compute the three-loop integral $\cA_3^{d=3}$ using the orbit method.
However, Eq. \eqref{Rhd} shows that the Rankin-Selberg transform is
given by 
\be
\label{R33}
\begin{split}
\cR^\star_3(\Gamma_{3,3,3};s) =& 
\zetastar(2s)\, \zetastar(2s-1)\,\zetastar(2s-2)\,\zetastar(2s-3)\, \\
& \times
\left[   
 \cE^{\star,SO(3,3)}_{S}(2s-1)+ \cE^{\star,SO(3,3)}_{C}(2s-1)
\right]\ .
\end{split}
\ee
The residue at $s=2$ produces 
\be
\cA_3^{d=3} = \zetastar(2) \,\zetastar(4) \,\left( \cE^{\star,SO(3,3)}_{S}(3)+ \cE^{\star,SO(3,3)}_{C}(3) \right)\ ,
\ee
in accordance with the conjecture in \cite{Obers:1999um}. The results for the three-loop amplitude
on $T^2$ and $S^1$ follow by  decompactification,
\be
\begin{split}
\cA_3^{d=2} =& \zetastar(2) \,\zetastar(4) \,\left( E_1(3;T) + E_1(3;U) \right)\\
\cA_3^{d=1} =& 2 \zetastar(2) \,\zetastar(4)\, \zetastar(6)\,\left( R^3 + 1/R^3 \right)
\end{split}
\ee
The $d=1$ result agrees with \eqref{Ah1}, it would be useful to obtain the $d=2$ case
from the subtraction $\delta$ in \eqref{Resdel}. 

\subsubsection{Modular integrals and spinor Eisenstein series}
While the relation \eqref{Rhd} between Langlands-Eisenstein series for the weights $\Lambda^h V,
S$ and $C$ only holds for $h=d$, it may be checked using Langlands' formula for the constant terms
that whenever $d\geq h$,
\be
\label{modspin}
{\rm Res}_{s=d/2} \cE^{SO(d,d),\star}_{\Lambda^h V}(s)-\delta= 
\frac18 r_h  \cV_h\, (1+\Theta_{d\leq 2h})\, 
{\rm Res}_{s=h} \left[  \cE^{SO(d,d),\star}_{S}(s) + \cE^{SO(d,d),\star}_{C}(s) \right]
\ee
where $\Theta_{d\leq 2h}=1$ if $d\leq 2h$ and 0 otherwise.
Hence, the genus $h$ modular integral ${\rm R.N.} \int_{\cF_h} \Gamma_{d,d,h}$ is proportional to a sum of (residues of) spinor Eisenstein series with $s=h$, 
as conjectured in  \cite{Obers:1999um}. It is worth noting that the residues of the two spinor
Eisenstein series appearing in \eqref{modspin} coincide only when $d\geq 2h+1$.

%\bibliography{autolec}
%\bibliographystyle{utphys}

\providecommand{\href}[2]{#2}\begingroup\raggedright\endgroup

\end{document}